\documentclass[a4paper,12pt,preprint]{aastex}
\usepackage{graphics,graphicx,rotating}

\newcommand{\tp}{\hspace{-1mm}+\hspace{-1mm}}
\newcommand{\tm}{\hspace{-1mm}-\hspace{-1mm}}
\newcommand{\cF}{{\cal F}}
\newcommand{\cG}{{\cal G}}
\newcommand{\cA}{{\cal A}}
\newcommand{\cD}{{\cal D}}
\newcommand{\dt}{\Delta\theta}
\newcommand{\db}{\Delta\beta}
\newcommand{\trQ}{{\rm tr}Q}
\newcommand{\paren}[1]{\left( #1 \right)}

\newcommand{\iotaIII}{\iota_{I\hspace{-.1em}I\hspace{-.1em}I}}
\newcommand{\upsilonII}{\upsilon_{I\hspace{-.1em}I}}
\newcommand{\upsilonIV}{\upsilon_{I\hspace{-.1em}V}}
\newcommand{\upsilonVI}{\upsilon_{V\hspace{-.1em}I}}
\newcommand{\tauIII}{\tau_{I\hspace{-.1em}I\hspace{-.1em}I}}
\newcommand{\tauVII}{\tau_{V\hspace{-.1em}I\hspace{-.1em}I}}
\newcommand{\Real}[1]{{\rm Re}\left[ #1 \right]}

\newcommand{\simgt}{\lower.5ex\hbox{$\; \buildrel > \over \sim \;$}}
\newcommand{\simlt}{\lower.5ex\hbox{$\; \buildrel < \over \sim \;$}}
\newcommand{\gIII}{I\hspace{-.3mm}I\hspace{-.3mm}I}

\def\bbeta{\mbox{\boldmath $\beta$}}
\def\btheta{\mbox{\boldmath $\theta$}}
\def\bnabla{\mbox{\boldmath $\nabla$}}

\def\bvtheta{\mbox{\boldmath $\vartheta$}}

\def\bphi{\mbox{\boldmath $\phi$}}

\shorttitle{A Method for Weak Lensing Flexion Analysis}
\shortauthors{Okura, Umetsu, \& Futamase}

\begin{document}


\title{A Method for Weak Lensing Flexion Analysis by the HOLICs Moment
 Approach\altaffilmark{1}}



\author{Yuki Okura\altaffilmark{2}} 
\email{aepgstrx@astr.tohoku.ac.jp}
\author{Keiichi Umetsu\altaffilmark{3}}
\email{keiichi@asiaa.sinica.edu.tw}
\and
\author{Toshifumi Futamase\altaffilmark{2}}
\email{tof@astr.tohoku.ac.jp}

\altaffiltext{1}{Based in part on data collected at the Subaru Telescope,
  which is operated by the National Astronomical Society of Japan}
\altaffiltext{2}
 {Astronomical Institute, Tohoku University, Sendai 980-8578, Japan}
\altaffiltext{3}
 {Institute of Astronomy and Astrophysics, Academia Sinica,  P.~O. Box 23-141, Taipei 106,  Taiwan, Republic of China}


\begin{abstract}
We have developed a method for measuring 
higher-order weak lensing distortions of faint background galaxies,
namely the weak gravitational flexion,
by fully extending the Kaiser, Squires \& Broadhurst method to include
higher-order lensing image characteristics (HOLICs) introduced by Okura,
 Umetsu, \& Futamase.
We take into account explicitly the weight function in
 calculations of noisy shape moments and the effect of 
higher-order
 PSF anisotropy, as well as isotropic PSF smearing.
Our HOLICs formalism allows accurate measurements of flexion from
practical observational data in the presence of non-circular,
anisotropic PSF.
We test our method using mock observations of simulated galaxy images
 and actual, ground-based
 Subaru observations of the massive galaxy cluster  A1689 
($z=0.183$).  From the high-precision 
measurements of spin-1 first flexion,
we obtain a high-resolution mass map in the central region
of A1689. 
The reconstructed mass map shows a bimodal feature in the central 
$4'\times 4'$ region of the cluster.
The major, pronounced peak is 
associated
 with the brightest cluster galaxy and central cluster members,
while the secondary mass peak is associated with a local concentration
of bright galaxies.
The refined, high-resolution mass map of A1689 
demonstrates the power of the generalized weak lensing analysis
techniques for quantitative and accurate measurements of the
weak gravitational lensing signal.
\end{abstract}






\section{Introduction}

Propagation of light rays from a distant source to the observer is
governed by the gravitational field of intervening mass fluctuations
as well as by the global geometry of the universe.
The images of background sources 
hence carry the imprint of the gravitational potential 
of intervening cosmic structures, and
their statistical properties
can be used to test the background cosmological models.

Weak gravitational lensing is responsible for the weak 
shape-distortion
and
magnification of the images of background sources due to the 
gravitational field of intervening matter
(e.g., Bartelmann \& Schneider 2001; Umetsu, Futamase, \& Tada 1999).
To the first order, weak lensing gives rise to 
a few -- $10\%$ levels of 
elliptical distortions
in images of background sources, responsible for the second-order
derivatives of the gravitational lensing potential.
Thus, the weak lensing signal, measured from tiny but coherent 
quadrupole distortions
in galaxy shapes, can provide a direct measure of the projected mass 
distribution of cosmic structures. 
However, practical weak lensing observations subject to
the effects of atmospheric seeing,
isotropic/anisotropic PSF, and (residual) camera distortion across the
field of view, 
which must be examined from the stellar shape measurements
and corrected for in the weak lensing analysis.
Practical methods for PSF corrections and shear measurements/calibrations
have been studied and developed by many authors, 
such as 
pioneering work of  Kaiser, Squires, \& Broadhurst (1995, hereafter
KSB), the Shapelets technique which describes PSF and object images in
terms of Gaussian-Hermite expansions (Refregier 2003),
and recent systematic, collaborative efforts by 
The Shear TEsting Programme  (Heymans et al. 2006; Massey et al. 2007).

Thanks to these successful developments in 
weak lensing techniques as well as in instrument technology,
the quadrupole weak lensing has become 
one of the most important tools in observational cosmology
to map the mass distribution 
in individual clusters of galaxies 
(e.g., 
Kaiser \& Squires 1993; 
Broadhurst et al. 2005a;
Okabe \& Umetsu 2007; 
Umetsu \& Broadhurst 2007),
measure ensemble-averaged mass profiles of galaxy-group sized halos
from the galaxy-galaxy lensing signal (e.g., Hoekstra et al. 2001; Hoekstra et
al. 2004; Parker et al. 2005; Mandelbaum et al. 2006), 
study the statistical properties of the 
large scale structure of the universe from the cosmic shear statistics
(e.g., Bacon et al. 2000;
van Waerbeke et al. 2001; Hamana et al. 2003), and search for
galaxy clusters by their mass properties 
(e.g., Schneider 1996; Erben et
al. 2000; Umetsu \& Futamase 2000; Wittman et al. 2001).

In recent years, there have been theoretical efforts to include
the next higher order distortion effects
as well as the usual quadrupole
distortion effect in the weak lensing analysis 
(Goldberg \& Natarajan 2002; 
Goldberg \& Bacon 2005; 
Bacon et al. 2006;
Irwin \& Shmakova 2006; 
Goldberg \& Leonard 2007; 
Okura, Umetsu, \& Futamase 2007).
We have proposed in Okura, Umetsu, \& Futamase (2007, hereafter OUF) 
to use
certain convenient combinations of octopole/higher multipole moments of
background images which we call the Higher Order Lensing Image's 
Characteristics (HOLICs), and have shown that
HOLICs serve as a direct measure for the next higher-order weak lensing
effect, or the gravitational flexion (Goldberg \& Bacon 2005)
and that the use of HOLICs in addition to the quadrupole shape
distortions can improve the accuracy and resolution of weak lensing
mass reconstructions based on simulated observations.

Recently, Goldberg \& Leonard (2007) 
extended the HOLICs approach for flexion measurements to
include observational effects, namely 
the Gaussian weighting in 
shape-moment calculations (see the appendix therein) and 
the isotropic PSF effect,
under the assumption that PSF is nearly circular,
and tested their extended HOLICs approach with simulated and HST/ACS
observations. Leonard et al. (2007) have applied the extended HOLICs
method to reconstruct the projected mass distribution in the central
region of the 
massive galaxy cluster A1689 at $z=0.183$, and revealed 
substructures associated with small clumps of galaxies.
Further
Leonard et al. (2007) found that in dense systems such as galaxy clusters
the HOLICs technique is robust and less sensitive than the Shapelet technique
to contamination by light
from the extended wings of lens/foreground galaxies.

In the present paper we develop a method for measuring flexion
by the HOLICs approach
by fully extending the KSB formalism;
We take into account explicitly the effects of 
Gaussian weighting in calculation of noisy shape moments and 
higher-order PSF anisotropy as well as isotropic PSF
smearing.
We then apply our method to actual, ground-based Subaru observations
of A1689, and perform a mass reconstruction in the central region of
A1689.

The paper is organized as follows. We first summarize in \S 2 the basis
of weak gravitational lensing and the flexion formalism.
In \S 3, we derive the relationship between HOLICs and flexion 
by incorporating Gaussian smoothing in shape measurements 
in the presence of isotropic and anisotropic PSF.
The practical method to correct the isotropic/anisotropic PSF effects
will be presented in \S 4. 
In Section 5 we use simulations to test our flexion analysis method
based on the HOLICs moment approach.
We then perform
a weak lensing flexion
analysis of A1689 by our fully-extended HOLICs approach, and perform a
mass reconstruction of A1689 from the HOLICs estimates of flexion.
Finally summary and discussions are given in \S 6.
We refer interested readers 
to a complete appendix\footnote
{Full appendix is available in electronic form at 
http://www.asiaa.sinica.edu.tw/keiichi/OUF2/appendix.pdf.}
for details of the derivation of flexion-observable relationships
in practical observations.

\section{Basis of Weak Lensing and Flexion}


In this section, 
we summarize general aspects of weak gravitational lensing and
flexion formalism, following the complex derivative notation
developed by Bacon et al. (2006).
A general review of quadrupole weak lensing can be found in 
Bartelmann \& Schneider (2001).

\subsection{Spin Properties}

We define the spin for weak-lensing quantities in the following way:
A quantity is said to have spin $N$ if it has the same value after 
rotation by $2\pi/N$. The product of spin-$A$ and spin-$B$ quantities
has spin $(A+B)$, and the product of spin-$A$ and spin-$B^*$ quantities
has spin $(A-B)$. Then, as we shall see in the next subsection,
the lensing convergence $\kappa$ is a spin-0 (scalar) quantity.
The complex shear $\gamma$ and the reduced shear $g=\gamma/(1-\kappa)$
are spin-2 quantities. The first and second flexion fields, $F$ and $G$,
are a spin-1 and a spin-3 quantity, respectively.

\subsection{Weak Lensing and Flexion Formalism}

The gravitational deflection of light rays can be described by the
lens equation,
\begin{equation}
\label{eq:lenseq}
\bbeta = \btheta - \bnabla \psi(\btheta),
\end{equation}
where $\psi(\btheta)$ is the effective lensing potential, which
is defined by the two-dimensional Poisson equation as
$\nabla^2\psi(\btheta)=2 \kappa(\btheta)$, with the lensing convergence.
Here the convergence $\kappa=\int\! d\Sigma_m \Sigma_{\rm crit}^{-1}$ is the
dimensionless surface mass density projected on the sky, normalized with
respect to the critical surface mass density of gravitational lensing,
\begin{equation}
\label{eq:sigma_crit}
\Sigma_{\rm crit} = \frac{c^2}{4\pi G}\frac{D_s}{D_d D_{ds}},
\end{equation}
where $D_d$, $D_s$, and $D_{ds}$ are the angular diameter distances
from the observer to the deflector, from the observer to the source,
and from the deflector to the source, respectively.
By introducing the complex gradient operator, $\partial = \partial_1 +
i\partial_2$ that transforms as a vector, 
$\partial'=\partial e^{i\phi}$, with $\phi$ being the angle of rotation,  
the lensing convergence $\kappa$ is expressed as
\begin{equation}
\label{eq:kappa}
\kappa = \frac{1}{2}\partial\partial^* \psi,
\end{equation}
where $^*$ denotes the complex conjugate.
Similarly, the complex gravitational shear of spin-2 is defined as
\begin{equation}
\gamma \equiv \gamma_1+i\gamma_2= \frac{1}{2}\partial\partial \psi.
\end{equation} 
The third-order derivatives of $\psi(\btheta)$ can be combined to
from a pair of the complex flexion fields as 
(Bacon et al. 2006):
\begin{eqnarray}
\label{eq:lpd3rd}
\cF &\equiv& \cF_1+i\cF_2 = \frac{1}{2}\partial\partial\partial^*\psi,\\
\cG &\equiv& \cG_1+i\cG_2 = \frac{1}{2}\partial\partial\partial\psi.
\end{eqnarray}
If the angular size of an image is small compared to the 
scale over which the 
lens  potential $\psi$ varies, then we can locally expand the lens
equation (\ref{eq:lenseq}) to have:
\begin{eqnarray}
\label{eq:dbetaij}
d\beta_i = \cA_{ij}d\theta_j + \frac{1}{2}\cD_{ijk}d\theta_j d\theta_k
\end{eqnarray}
to the second order, where $\cA_{ij}$ is the Jacobian matrix of the lens
equation and $\cD_{ijk}=\cA_{ij,k}=-\psi_{,ijk}$ is the third-order
lensing tensor,
\begin{eqnarray}
\cA_{ij}&=&
\left( 
\begin{array}{cc} 
1-{\kappa}-{\gamma}_1 & -{\gamma}_2 \\
 -{\gamma}_2  & 1-{\kappa}+{\gamma}_1 
\end{array}
\right),\\
\cD_{ijk}&=&\cF_{ijk}+\cG_{ijk}.
\end{eqnarray}
The third-order tensor $\cD_{ijk}$ can be expressed with the sum of the
two terms, $\cD_{ijk}=\cF_{ijk}+\cG_{ijk}$, with the spin-1 part
$\cF_{ijk}$ and the spin-3 part $\cG_{ijk}$, composed of the
real/imaginary part of the flexion fields:
\begin{eqnarray}
	{\cal F}_{ij1} = -\frac{1}{2}\left(
\begin{array}{@{\,}cc@{\,}}
		3{\cal F}_1  & {\cal F}_2 \\
		 {\cal F}_2  & {\cal F}_1 
	\end{array}
\right) &,& \ \ 
{\cal F}_{ij2} = -\frac{1}{2}\left(
\begin{array}{@{\,}cc@{\,}}
		{\cal F}_2  &  {\cal F}_1 \\
		{\cal F}_1  & 3{\cal F}_2
	\end{array}
\right), \\
{\cal G}_{ij1} = -\frac{1}{2}\left(
\begin{array}{@{\,}cc@{\,}}
		{\cal G}_1 &  {\cal G}_2 \\
		{\cal G}_2 & -{\cal G}_1
	\end{array}
\right)&,& \ \ 
{\cal G}_{ij2} = -\frac{1}{2}\left(
\begin{array}{@{\,}cc@{\,}}
		 {\cal G}_2 & -{\cal G}_1 \\
		-{\cal G}_1 & -{\cal G}_2 
	\end{array}
\right).
\end{eqnarray}
Note that flexion has a dimension of inverse length (or inverse angle),
meaning that the flexion effect depends on the angular size of the
source.
The shape quantities affected by the first flexion $\cF$ alone have
spin-1 properties, while those affected by the second flexion $\cG$
alone have spin-3 properties. These third-order lensing fields naturally
appear in the transformation equations of HOLICs between the lens and
source planes.

\subsection{Quadrupole Lensing Observable -- Complex Ellipticity}
 

In the KSB approach, we use quadrupole moments $Q_{ij}$
of the surface brightness distribution $I(\btheta)$ of background images
for quantifying the shape of the images:
\begin{equation}
\label{eq:Qij}
Q_{ij} \equiv \frac{\int\! d^2\theta\, 
q_I[I(\btheta)]\dt_i \dt_j}{\int d^2\theta\,q_I[I(\btheta)]},
\end{equation}
where $q_I[I(\btheta)]$ denotes the weight function used in 
noisy shape measurements and $\Delta\theta_i = \theta_i-\bar{\theta}_i$
is the offset vector from the image centroid.
The complex ellipticity $\chi$ is then defined as
\begin{equation}
\label{eq:cellip}
\chi \equiv \frac{Q_{11} - Q_{22} + 2iQ_{12}}{Q_{11} + Q_{22}},
\end{equation}
The $\chi$ transforms under the lens mapping as
\begin{equation}
\label{eq:chis2chi}
\chi^{(s)}=\frac{\chi-2g+g^2\chi^\ast}{1+|g|^2-2\Real{g\chi^\ast}},
\end{equation}
where $g = \gamma/(1 - \kappa)$ is the spin-2 reduced shear.
In the weak lensing limit ($\kappa, |\gamma|\ll 1$),
equation (\ref{eq:chis2chi}) reduces to 
$\chi^{(s)} \approx \chi-2\gamma$.
Assuming the random orientation of the background sources,
we average observed ellipticities over a sufficient number of
images to obtain
\begin{equation}
\label{eq:chis2chiap}
\langle\chi\rangle\approx 2g \approx 2\gamma.
\end{equation}

\subsection{Centroid Shift due to Lensing}
\label{subsec:csfgl}

In the moment methods such as KSB, we quantify the shape of
an image by measuring various moments of $I(\btheta)$, in which
the moments are calculated with respect to the observable
centroid of the image $\bar{\btheta}$, 
or the center of the light, defined by the
first moment of $I(\btheta)$.
However, in the presence of gravitational lensing, this apparent center
can be different from the point $\btheta(\bar{\bbeta})$
that is mapped using the lens
equation (\ref{eq:lenseq})
from the center of the unlensed light, $\bar{\bbeta}$. 
We refer to this point
as the ``true'' center of the image. The difference between these two
centers, namely, the apparent and the true centers, causes a significant
effect in evaluating the first flexion, as pointed out by Goldberg \&
Bacon (2005). The relationship between the two centers is 
\begin{equation}
\label{eq:csft}
\theta_i(\bar{\bbeta}) 
\approx  \bar{\theta}_i - \trQ 
\paren{\frac{3}{2} F_i + \frac{5}{4}[F^*\chi]_i +
\frac{1}{4}[G\chi^*]_i}
\equiv \bar{\theta}_i - \Delta_{L,i},
\end{equation}
where $\trQ=Q_{11}+Q_{22}$ is the trace of $Q_{ij}$, 
$F$ and $G$ are the 
reduce Flexion, defined by 
$F=\cF/(1-\kappa)$ and $G=\cG/(1-\kappa)$, respectively,
and $\Delta_{L,i}$ is the displacement vector from the true to the
apparent center due to gravitational lensing.
Therefore, by taking into account the centroid shift in shape 
measurements,
we have the relation between $\Delta\beta_i=\beta_i-\bar{\beta}_i$ and 
$\Delta\theta_i = \theta_i -\bar{\theta}_i$ as 
\begin{eqnarray}
\Delta\beta_i &\approx& 
\cA_{ij}\Delta\theta_j 
+ \frac{1}{2}\cD_{ijk} \Delta\theta_j \Delta\theta_k 
+ \trQ \paren{\frac{3}{2} F_i + \frac{5}{4}[F^*\chi]_i 
+ \frac{1}{4}[G\chi^*]_i}\nonumber\\
&=& 
\cA_{ij}\Delta{\theta}_j + \frac{1}{2}\cD_{ijk}\Delta\theta_j
\Delta\theta_k 
+ \Delta_{L,i}.
\end{eqnarray}
A detailed derivation of the above relation is given in 
Appendix B of OUF.

\subsection{Flexion Observable -- HOLICs}

The flexion fields, $\cF$ and $\cG$, can be measured from 
proper combinations of higher-order shape moments with the
corresponding spin properties and the dimension, 
as explicitly shown by OUF.
Higher-order moments of images are defined as a straightforward
extension of the quadrupole moment.
The octopole moment $Q_{ijk}$
 and the 16pole moment $Q_{ijkl}$ are defined as follows:
\begin{eqnarray}
\label{eq:Qijk}
Q_{ijk}&\equiv& \frac{\int\! d^2\theta\, q_I[I(\btheta)]\dt_i \dt_j \dt_k}{\int d^2\theta\,q_I[I(\btheta)]} \\
Q_{ijkl}&\equiv& \frac{\int\! d^2\theta\, q_I[I(\btheta)]\dt_i \dt_j \dt_k \dt_l}{\int d^2\theta\,q_I[I(\btheta)]}.
\end{eqnarray}
Then, $\zeta$ and $\delta$ of the spin-1 and spin-3 HOLICs,
 respectively,
 are defined by
\begin{eqnarray}
\label{eq:HOLICs}
\zeta&\equiv&\frac{Q_{111}+Q_{122}+i\left(Q_{112}+Q_{222}\right)}{\xi}\nonumber\\
\delta&\equiv&\frac{Q_{111}-3Q_{122}+i\left(3Q_{112}-Q_{222}\right)}{\xi},
\end{eqnarray}
where $\xi$ is the spin-0 normalization factor,
\begin{equation}
\xi = Q_{1111}+2Q_{1122}+Q_{2222}.
\end{equation}

Finally, the transformation equations between unlensed and lensed HOLICs
are obtained as (see OUF)
\begin{eqnarray}
\label{eq:zeta}
\zeta^{(s)}
&=&
\frac{\zeta-2g\zeta^*-g^*\delta -\frac{1}{4}(8F^*\eta - 16\frac{\paren{\trQ}^2}{\xi}F^*\chi +9F - 12\frac{\paren{\trQ}^2}{\xi}F +2G\eta^* - 2\frac{\paren{\trQ}^2}{\xi}G\chi^*+G^*\lambda)}{(1-\kappa)(1-4\Real{g^*\eta}-5\Real{F\iota_I^*}-\Real{G\iotaIII^*})},\nonumber\\
\label{eq:delta}
\delta^{(s)}
&=
&\frac{\delta-3g\zeta -\frac{1}{4}(10F\eta+7F^*\lambda - 18\frac{\paren{\trQ}^2}{\xi}F\chi +3G)}{(1-\kappa)(1-4\Real{g^*\eta}-5\Real{F\iota_I^*}-\Real{G\iotaIII^*})},
\end{eqnarray}
where $\eta$ and $\lambda$ are dimensionless spin-2 and spin-4
quantities, respectively,
defined with 16-pole moments,
and $\iota_{I}$, $\iota_{\gIII}$ and $\iota_{V}$ are 
spin-1, spin-3, and spin-5 quantities, respectively,
defined with $32$-pole moments (see Appendix and OUF).
We note that, the above equations (\ref{eq:zeta}) and (\ref{eq:delta})
are obtained under the sub-critical lensing condition, i.e., ${\rm
det}{\cal A}>0$ (see Schneider \& Er 2007).

Since the HOLICs $\zeta$ and $\delta$ are non-zero spin quantities
with a direction dependence, the expectation value of the intrinsic
$\zeta$ and $\delta$ are assumed to vanish. To the first order in
flexion,
we have the linear relations between the HOLICs and flexion fields as
\begin{eqnarray}
\label{eq:HOLICFL}
F &\approx& 
\left<\frac{\zeta}{\frac{9}{4} - 3\frac{\paren{\trQ}^2}{\xi}}\right>\\
G &\approx& \frac{4}{3}\left<\delta\right>.
\end{eqnarray}

\section{HOLICs and Flexion in Practical Applications}
\label{sec:ksb+}

For a practical application of the HOLICs approach,
we must take into account various observational effects
such as noise in the shape measurement due to readout
and/or sky background
and the dilution of the lensing signal due to the isotropic/anisotropic
PSF effects.
Thus, one cannot simply use equations 
(\ref{eq:zeta}) and (\ref{eq:delta})
to measure the flexion fields.
In this section, we introduce the Gaussian weighting in moment
calculations, as done in the KSB formalism for quadrupole weak lensing, 
and derive the relevant
transformation equations between unlensed and lensed HOLICs
by taking into account
explicitly the effect of Gaussian smoothing.

\subsection{Redefining HOLICs for Noisy Observations}

Now we introduce a weight function $W(|\dt|^2/\sigma^2)$
having a characteristic width $\sigma$ for practical, noisy
moment measurements,
redefining the octopole and 16-pole moments of the 
brightness distribution $I(\btheta)$
as
\begin{eqnarray}
\label{octW}
Q_{ijk}&\equiv&
  \int d^2\theta I(\theta)\dt_i\dt_j\dt_kW(|\dt|^2/\sigma^2),\\
Q_{ijkl}&\equiv&
  \int d^2\theta I(\theta)\dt_i\dt_j\dt_k\dt_lW(|\dt|^2/\sigma^2),
\end{eqnarray}
where $Q$s here are no longer normalized with the corresponding {\it flux}
of the image (see equation [\ref{eq:Qij}]).
The redefined shape moments enter equation (\ref{eq:HOLICs}). 
We provide in Appendix
detailed definitions of HOLICs in practical applications.

\subsection{Lensing-Induced Centroid Shift in Weighted Moment Calculations}

When weighted moments are used for calculating HOLICs,
the centroid shift due to lensing (\ref{eq:csft}), to the first order,
is changed in the following way:
\begin{eqnarray}
\label{SCL}
\Delta_{L} = \bar{\theta} - \theta(\bar{\beta})
\approx 
\frac{\frac{3\trQ^a}{2M} + \frac{3{\xi^a}'}{4M\sigma^2}}{1+\frac{{\trQ^a}'}
{M\sigma^2}}F \equiv \Delta^0_L F,
\end{eqnarray}
where 
$\bar{\theta} = \bar{\theta}_1+i\bar{\theta}_2$ 
and $\theta(\bar\beta) = \theta_1(\bar\beta)+i\theta_2(\bar\beta)$ 
are the apparent and the true centers of the image
(see \S \ref{subsec:csfgl}), respectively, 
in the complex form calculated using the weight function $W(x)$, 
$M\equiv \int\!d^2\theta\,I(\theta)W(|\Delta\theta|^2/\sigma^2)$ 
is the monopole shape moment (or {\it flux}),
quantities with subscript ``a'' represent those calculated
with respect to the apparent center,
and quantities with prime represent those measured with 
$W'(x)=\partial W(x)/\partial x$ as the weight function 
instead of $W(x)$; $\Delta_L^0$ is the spin-0 coefficient in $\Delta_L$.
The deviation from unity in 
the denominator of $\Delta_L^0$ is obtained by properly expanding
the weight function W(x) in moment calculations, and this term does not
appear in Goldberg \& Leonard (2006)'s formulation.
Hence,
the complex displacement from the true image center, $\dt^t$,
can be expressed in terms of that from the apparent image center,
$\dt^a$, and the complex centroid shift, $\Delta_L$, as
\begin{eqnarray}
\label{eq:csf_complex}
\dt^t = \dt^a + \Delta_L.
\end{eqnarray}
For interested readers, we refer to Full Appendix B.1 for detailed
calculations of the 
lensing-induced centroid shift with a weight function.

\subsection{Relation between the Weighted HOLICs and Flexion}


In weighted moment calculations, the transformation equations
between HOLICs and flexion must be modified accordingly.
To the first order, we have
\begin{eqnarray}
\label{eq:Lbt}
\zeta^{(s)}
&\approx&
\frac{1}{1-\kappa}\left(\zeta^t-\frac{9}{4}F-\frac{3{\upsilon^t_0}'}
{4\xi^t\sigma^2}F\right),\\
\delta^{(s)}&\approx&\frac{1}{1-\kappa}\left(\delta^t-
\frac{3}{4}G-\frac{{\upsilon^t_0}'}{4\xi^t\sigma^2}G\right),\\
\upsilon_0&\equiv&Q_{111111}+3Q_{112222}+3Q_{111122}+Q_{222222},
\end{eqnarray}
where quantities with subscript ``t'' represent those calculated 
with respect to the true center, or using $\Delta\theta^t$.
By using equation (\ref{eq:csf_complex}),
we can express $\zeta^t$ and $\delta^t$ in terms of 
practically observable quantities 
(with subscript ``a'')
as
\begin{eqnarray}
\label{eq:zeta_t}
\zeta^t &\approx&
\zeta^a + 2\frac{\trQ^a}{\xi^a}\Delta_L+\frac{{\xi^a}'}{\xi^a\sigma^2}\Delta_L
= \zeta^a + \left(2\frac{\trQ^a}{\xi^a}\Delta^0_L+\frac{{\xi^a}'}
{\xi^a \sigma^2}\Delta^0_L\right)F,\\
\label{eq:delta_t}
\delta^t&\approx&\delta^a
\end{eqnarray}
to the first order.
Here the term $\frac{{\xi^a}'}{\xi^a\sigma^2}\Delta_L$ 
in equation (\ref{eq:zeta_t}) is again caused by the centroid shift
in the weight function 
(a similar, but different, expression was
obtained by Goldberg \& Leonard 2007).
Finally, we obtain the following transformation equations
in the case of weighted moment calculations:
\begin{eqnarray}
\label{eq:Lbf}
\zeta^{(s)}&
\approx& 
\frac{1}{1-\kappa}
\left[
\zeta^a -
\left(\frac{9}{4}+\frac{3{\upsilon^a_0}'}{4\xi^a\sigma^2}-
2\frac{\trQ^a}{\xi^a}\Delta^0_L-\frac{{\xi^a}'}{\xi^a\sigma^2}\Delta^0_L
\right)
F
\right],\\
\delta^{(s)}
&\approx&
\frac{1}{1-\kappa}
\left[
\delta^a-\left(\frac{3}{4}+\frac{{\upsilon^a_0}'}{4\xi^a\sigma^2}\right)G
\right]
\end{eqnarray}
We show the details of these calculations in Full Appendix B.3.

\section{Isotropic and Anisotropic PSF Corrections for HOLICs Measurements}

In this section we present a detailed prescription for
the PSF anisotropy and circularization correction in HOLICs-based
flexion measurements by extending the KSB formalism. 
We closely follow the treatment and the notation
given in \S 4.6.1 of Bartelmann \& Schneider (2001).

\subsection{General Description for PSF}

The observed surface brightness distribution $I^{\rm obs}(\btheta)$
can be expressed as the true surface brightness $I(\btheta)$
convolved with an effective PSF $P(\btheta)$, 
\begin{eqnarray}
\label{eq:PSF}
I^{\rm obs}(\btheta) 
= \int\! d^2\vartheta\, 
I(\bvtheta)P(\btheta-\bvtheta).
\end{eqnarray}
Following the KSB formalism,
we assume that $P$ is nearly isotropic, so that the 
anisotropic part of $P$ is small. We then define the 
isotropic part of $P$ as the azimuthal
average over $P$, and decompose $P$ 
into an isotropic part, $P^{iso}$, 
and an anisotropic part, $q$,
 as
\begin{eqnarray}
\label{eq:PSFP}
P(\bvtheta) 
= \int\! d^2\phi\, q(\bphi) P^{iso}(\bvtheta - \bphi),
\end{eqnarray}
where both $P^{iso}$ and $q$ are normalized to unity.
We then define $I^{iso}(\btheta)$ as the surface brightness distribution
smeared by the isotropic part $P^{iso}$,
\begin{eqnarray}
\label{eq:PSFiso}
I^{iso}(\btheta) 
= \int\! d^2\vartheta\, I(\btheta)P^{iso}(\btheta-\bvtheta).
\end{eqnarray}
The observed surface brightness $I^{obs}(\btheta)$ is obtained by
convolving $I^{iso}(\btheta)$ with the anisotropy kernel $q(\btheta)$
as
\begin{eqnarray}
\label{eq:PSFaniso}
I^{obs}(\btheta) = \int\! d^2\vartheta\, 
q(\btheta - \bvtheta)I^{iso}(\bvtheta).
\end{eqnarray}

In the original KSB method for quadrupole weak lensing,
only the spin-2 PSF anisotropy described by
the quadrupole moment of the anisotropy kernel $q$
has to be taken into account.
However, 
in order to correct HOLICs with spin-1 and spin-3 properties
for the anisotropic PSF effects,
we need to take into account the corresponding
dipole and octopole moments of the anisotropy kernel $q$
having spin-1 and spin-3 properties.
We expand the integral of arbitrary function $f(\btheta)$ 
and $I^{obs}(\btheta)$ to obtain
\begin{eqnarray}
\label{eq:psfexp}
&&\int\! d^2\theta\,
 f(\btheta) I^{obs}(\btheta) = 
\int\! d^2\phi\, I^{iso}(\bphi) 
\int\! d^2\theta\, f(\btheta +\bphi) q(\btheta)
\approx \int\! d^2\phi\,
  I^{iso}(\bphi) f(\bphi)\nonumber\\
&&\hspace{1cm} 
+ q_{k}\int\! d^2\phi\,
   I^{iso}(\bphi) 
    \frac{\partial f}{\partial\phi_k} 
  + \frac{1}{2}q_{kl}
 \int\! d^2\phi\, 
  I^{iso}(\bphi) \frac{\partial^2f}{\partial\phi_k\partial\phi_l}\nonumber\\
&&\hspace{1cm}
+ 
    \frac{1}{6} q_{klm}
 \int\! d^2\phi\, 
  I^{iso}(\bphi)\frac{\partial^3f}{\partial\phi_k\partial\phi_l\partial\phi_m}
 + .........,
\end{eqnarray}
where
\begin{eqnarray}
\label{eq:qijk}
&q_{i} =   \int\! d^2\phi\, q(\bphi)\phi_i,\\
&q_{ij} =  \int\! d^2\phi\, q(\bphi)\phi_i\phi_j,\\
&q_{ijk} = \int\! d^2\phi\, q(\bphi)\phi_i\phi_j\phi_k
\end{eqnarray}
are the dipole, quadrupole, and octopole moments of the 
PSF anisotropy kernel $q$, respectively.

\subsection{Centroid Shift due to PSF Anisotropy}

The PSF anisotropy can cause a centroid shift between the 
images defined in terms of 
$I^{iso}(\btheta)$ and $I^{obs}(\btheta)$.
This centroid shift is essentially due to the
spin-1 PSF anisotropy.

The observed weighted-center of image can be expressed 
in the complex form as
\begin{equation}
\label{eq:Cb_01}
\bar \theta^{obs} = \frac{\int\! d^2\theta\, \theta I^{obs}(\theta) 
W(|\theta|^2/\sigma^2)}
{\int\! d^2\theta\, I^{obs}(\theta)W(|\theta|^2/\sigma^2)},
\end{equation}
where $\theta=\theta_1+i\theta_2$ is the complex angular position.
Similarly, 
the weighted-center of a hypothetical image defined in terms of
$I^{iso}$ is 
\begin{equation}
\label{eq:Cb_02}
\bar \theta^{iso}=\frac{\int\! d^2\theta\,\theta I^{iso}(\theta)
W(|\theta|^2/\sigma^2)}{\int\! d^2\theta\,  I^{iso}(\theta) 
W(|\theta|^2/\sigma^2)}.
\end{equation}
Then, the offset between the two centers $\bar{\theta}^{obs}$ and
$\bar{\theta}^{iso}$ is given as 
\begin{eqnarray}
\Delta_P \equiv\bar \theta^{obs} - \bar\theta^{iso}
= \dt^{iso}-\dt^{obs},
\end{eqnarray}
where $\dt^{iso} \equiv \theta-\bar{\theta}^{iso}$ and 
$\dt^{obs} \equiv \theta-\bar{\theta}^{obs}$ are the complex displacements
to an arbitrary point $\theta$.
Expanding $\Delta_P$ to the first order of $q$ yields
\begin{eqnarray}
\Delta_P
= \Delta\theta^{iso} - \Delta\theta^{obs}
\approx D_q + \frac{\frac{M}{M'\sigma^2} 
+ \frac{2\trQ''}{M\sigma^4}
+ \frac{\xi'''}{2\sigma^6 M}}{1+\frac{\trQ'}{M\sigma^2}}\zeta_q 
\equiv D_q + \frac{P^0_D}{P^\Delta_D}\zeta_q,
\end{eqnarray}
where
quantities with $\prime\prime$ and $\prime\prime\prime$
refer to those calculated with
$W''(x)$ and $W^{\prime\prime\prime}(x)$ 
as the weight function, respectively,
$(D_q,\zeta_q)$ are 
spin-1 complex moments of the anisotropy kernel defined by
\begin{eqnarray}
D_q &=& \int d^2\phi q(\phi) \phi,\\
\zeta_q &=& \int d^2\phi q(\phi) \phi\phi\phi^*,\\
\phi&=&\phi_1+i\phi_2,
\end{eqnarray}
and $(P^0_D,P^\Delta_D)$ are
certain combinations of
shape moments defined as the coefficients 
in equation (C11) of Full Appendix C.1.2,
associated with the complex dipole moment $D$ with spin-1 properties.
We note that $\Delta_P$, the PSF-induced centroid shift, cannot be
constrained from observations; however, as we shall see in the next
subsection, one can constrain the spin-1 octopole component in
$\Delta_P$, namely $\zeta_q \propto \Delta_P-D_q$,
in the first order approximation.
More detailed calculations are presented in Full Appendix C.1.2.

\subsection{Anisotropic PSF Correction for HOLICs}

We use equation (\ref{eq:psfexp}) to relate
observable HOLICs (with subscript ``obs'')
to those defined in terms of $I^{iso}$ (with subscript ``iso'').
To the first order of $q$, we have the following relations
for the spin-1 and spin-3 quantities:
\begin{eqnarray}
\label{eq:HOLICisoz}
D^{obs}
&\approx&
D^{iso} + \Delta_P
\approx
D^{iso} + D_q + \frac{P^0_D}{P^\Delta_D}\zeta_q,\\
\zeta^{obs}
&\approx&
\zeta^{iso} + \frac{1}{\xi}\left( 2\trQ 
   + \frac{\xi'}{\sigma^2} \right)(D_q-\Delta_P) 
   + \frac{1}{\xi}\left(M+\frac{5\trQ'}{\sigma^2}
   + \frac{7\xi''}{2\sigma^4}+\frac{\upsilon'''_0}{2\sigma^6}\right)
  \zeta_q\nonumber\\
  &=& \zeta^{iso} 
  + \left( P^0_\zeta - \frac{P^0_D}{P^\Delta_D}P^\Delta_\zeta 
    \right)\zeta_q,\\
\label{eq:HOLICisod}
\delta^{obs}
&\approx&
\delta^{iso}
  +\frac{1}{\xi}\left(M+\frac{3\trQ'}{\sigma^2}
  +\frac{3\xi''}{2\sigma^4}+\frac{\upsilon'''_0}{6\sigma^6}\right)
 \delta_q=\delta^{iso}+P^0_{\delta}\delta_q,
\end{eqnarray}
where we have deffined the weighted first moment $D$ by
\begin{equation}
D=\frac{\int\!d^2\theta\, W(|\Delta\theta|^2/\sigma^2)\Delta\theta}{M},
\end{equation}
and the spin-3 PSF anisotropy $\delta_q$ by
\begin{equation}
\delta_q = \int\! d^2\theta\, q(\phi)\phi\phi\phi.
\end{equation}
In general, the PSF varies spatially over the field.
If the spatial variation of PSF is sufficiently smooth,
then one can measure $\zeta_q$ and $\delta_q$ for a set of stars.
Since $\zeta^{iso}$ and $\delta^{iso}$ vanish for stars,
the spin-1 and spin-3 PSF anisotropies can be obtained as
\begin{eqnarray}
\label{eq:qstar}
\zeta_q&=&\frac{   
(\zeta^{obs})_{*}
}
{\left( P^0_\zeta - \frac{P^0_D}{P^\Delta_D}P^\Delta_\zeta 
\right)_{*}
},\\
\delta_q &=&
\frac{
(\delta^{obs})_{*}
}
{ (P^0_{\delta})_{*}},
\end{eqnarray}
where quantities with asterisk denote those measured for stellar objects.
Note that unlike the higher-order PSF anisotropies $\zeta_q$ 
and $\delta_q$, $\Delta_p$ cannot be determined from observations,
so that $D^{obs}$ cannot be corrected for the anisotropic PSF effect.
We show detailed calculations in Full Appendix C.1.3.

\subsection{Isotropic PSF Correction for HOLICs}

This subsection provides the method of correcting for the 
isotropic PSF effect on the flexion measurement.
Firstly, from Liouvelle's theorem we have 
$I(\btheta)=I^{(s)}(\bbeta(\btheta))$ with $I^{(s)}(\bbeta)$
being the surface brightness distribution of the unlensed source.
We then consider  
\begin{eqnarray}
I^{iso}(\btheta) &=& \int\!d^2\phi\,
 I^{(s)}(\cA \bphi)P^{iso}(\btheta-\bphi)\nonumber\\
&=& \frac{1}{{\rm det}\cA}\int\!d^2\vartheta\, I^{(s)}(\bvtheta)
 P^{iso}(\btheta-\cA^{-1}\bvtheta) \equiv \hat{I}(\cA\btheta),
\end{eqnarray}
where we have defined the brightness distribution $\hat{I}(\bbeta)$
convolved with the hypothetical PSF $\hat{P}(\bbeta)$,
\begin{eqnarray}
\hat{I}(\bbeta) &=& \int\!d^2\phi\, I^{(s)}(\bphi)\hat{P}(\bbeta-\bphi),\\
\hat{P}(\bbeta) &=& \frac{1}{{\rm det}\cA} P^{iso}(\cA^{-1}\bbeta).
\end{eqnarray}
The $\hat{P}$ can be regarded as an effective PSF relating 
$\hat{I}$ to
$I^{(s)}$; the anisotropic part of $\hat{P}$ is caused by lensing.
The octopole moment defined with $\hat I(\beta)$ is written as
\begin{eqnarray}
\hat Q_{ijk} =\int\! d^2\beta\, 
\Delta\beta_i 
\Delta\beta_j 
\Delta\beta_k 
\hat I(\bbeta) 
W\paren{\frac{|\Delta\bbeta|^2}{\hat \sigma^2}}.
\end{eqnarray}
Then, $\hat \zeta$ and $\hat \delta$ of HOLICs are defined 
in terms of $\hat{Q}_{ijk}$ and $\hat{Q}_{ijkl}$
as
\begin{eqnarray}
\hat \zeta 
&=&\frac{\hat Q_{111}+\hat Q_{122}+i(\hat Q_{112}+\hat Q_{222})}
{\hat Q_{1111}+2\hat Q_{1122}+\hat Q_{2222}},\\
\hat \delta &=&\frac{\hat Q_{111}-3\hat Q_{122}+i(3\hat Q_{112}-\hat
 Q_{222})}
{\hat Q_{1111}+2\hat Q_{1122}+\hat Q_{2222}}.
\end{eqnarray}
Substituting the expression for $\db$ 
into the above equations and using equation (\ref{eq:Lbf}),
we obtain
\begin{eqnarray}
\label{hatiso}
\hat \zeta 
&\approx&
\zeta^{iso}- \left(\frac{9}{4}
 +\frac{3{\upsilon_0^{iso}}'}{4\xi^{iso}}
 -2\frac{\trQ^{iso}}{\xi^{iso}}\Delta^0_L
 -\frac{{\xi^{iso}}'}{\xi^{iso}\sigma^2}\Delta^0_L\right)F
=\zeta^{iso} - 
\left(C^0_{\zeta}+C^\Delta_\zeta\Delta^0_L\right) F,\\
\hat \delta &\approx&
\delta^{iso}-\left(\frac{3}{4}
+\frac{{\upsilon_0^{iso}}'}{4\xi^{iso}}\right)G 
= \delta^{iso} - C^0_\delta G
\end{eqnarray}
to the first order.
Here $C^0$s are certain combinations of
shape moments defined in terms of $I^{iso}$ 
(see Full Appendix C.2).
In practice, however, 
one can replace $I^{iso}$ with $I^{obs}$ for calculating $C^0$s
of first order in flexion. 

Next, we
decompose 
$\hat P(\bbeta)$ into an isotropic part, $\hat P^{iso}$, 
and an anisotropic part, $\hat q$, as
\begin{eqnarray}
\hat P(\bbeta) = 
\int\! d^2\phi\, \hat q(\bbeta - \bphi)\hat P^{iso}(\bphi).
\end{eqnarray}
With $\hat P^{iso}$ we define the surface brightness distribution
$\hat I^0$ smeared by the isotropic part $\hat P^{iso}$
as
\begin{eqnarray}
\hat I^0(\bbeta) = \int\! d^2\phi\, 
I^{(s)}(\bphi)\hat P^{iso}(\bbeta - \bphi).
\end{eqnarray}
Then, the relationship between $\hat I(\bbeta)$ and $\hat I^0(\bbeta)$
is the same as that between $I^{obs}(\btheta)$ 
and $I^{iso}(\btheta)$
except that $\hat P$ instead of $P$.
Therefore, we obtain the following relations to the first order:
\begin{eqnarray}
\label{hat0}
\hat \zeta 
 &\approx&\hat \zeta^0 + \hat P^0_\zeta\zeta_{\hat q},\\
\hat \delta &\approx&\hat \delta^0 + \hat P^0_\delta\delta_{\hat q},
\end{eqnarray}
where $\zeta_{\hat q}$ and $\delta_{\hat q}$ are defined as
\begin{eqnarray}
\zeta_{\hat q}  &=& \int\!d^2\theta\, \hat q(\phi)\phi\phi\phi^*,\\
\delta_{\hat q} &=& \int\!d^2\theta\, \hat q(\phi)\phi\phi\phi,
\end{eqnarray}
$\hat P^0_\zeta$  and $\hat P^0_\delta$
are spin-0 coefficients defined by certain combinations of shape moments
(see Full Appendix C.2), 
and are calculated using $I^0$ instead
of $I^{iso}$ to the first order approximation.
For stellar objects, $\hat\zeta_*^0=\hat\delta_*^0=0$,
so that higher-order PSF anisotropies of $\hat q$ are obtained
using equations  (\ref{hatiso}) and 
(\ref{hat0}) as
\begin{eqnarray}
\zeta_{\hat q}&=&
-\frac{\left(C^0_{\zeta}+C^\Delta_\zeta\Delta^0_L\right)_{*}}
{(1-\kappa)\left( \hat P^0_\zeta 
- \frac{\hat P^0_D}{\hat P^\Delta_D}\hat P^\Delta_\zeta \right)_{*}}F,\\
\delta_{\hat q} &=&
-\frac{  (C^0_{\delta})_* }{(1-\kappa) (\hat P^0_{\delta})_* }G.
\end{eqnarray}
Further, to the first order approximation,
we have
\begin{eqnarray}
\left( 
 \hat P^0_\zeta - \frac{\hat P^0_D}{\hat P^\Delta_D}\hat P^\Delta_\zeta 
\right) 
&\approx&
(1-\kappa)^{-3}
\left( 
P^0_\zeta - \frac{P^0_D}{P^\Delta_D}P^\Delta_\zeta
\right),
\\
\hat P^0_\delta&\approx&\frac{P^0_\delta}{(1-\kappa)^3}
\end{eqnarray}
Finally,
we can relate 
unlensed HOLICs in terms of $\hat I^0(\bbeta)$ to 
observed HOLICs corrected for the PSF anisotropy
by
\begin{eqnarray}
\label{iso0}
\hat \zeta^0 &\approx&
\frac{1}{1-\kappa}
\left[
 \zeta^{iso}- \left(C^0_{\zeta}+C^\Delta_\zeta\Delta^0_L\right) F
 +
 \frac{\left(C^0_{\zeta}+C^\Delta_\zeta\Delta^0_L\right)_{*}}
      {\left(P^0_\zeta
        -
       \frac{P^0_D}{P^\Delta_D}P^\Delta_\zeta\right)_{*}}
   \left(P^0_\zeta - \frac{P^0_D}{P^\Delta_D}P^\Delta_\zeta\right) F
\right],
\\
\hat \delta^0 &\approx&
\frac{1}{1-\kappa}
\left[
\delta^{iso}- C^0_\delta G + \frac{ (C^0_{\delta})_* }
 { (P^0_{\delta})_* }P^0_\delta G
\right].
\end{eqnarray}
Assuming $\langle \hat\zeta^0 \rangle = \langle \hat\delta^0\rangle =0 $
for unlensed sources,
we obtain the desired expressions for flexion as
\begin{eqnarray}
\label{eq:HtoF}
F &\approx& 
\left\langle 
   \frac{\zeta^{iso}}{
       \left(C^0_{\zeta} + C^\Delta_\zeta\Delta^0_L\right)
         -
      \frac{\left(C^0_{\zeta}+C^\Delta_\zeta\Delta^0_L\right)_{*}}
   {\left(P^0_\zeta -
     \frac{P^0_D}{P^\Delta_D}P^\Delta_\zeta\right)_{*}}
   \left(P^0_\zeta - \frac{P^0_D}{P^\Delta_D}P^\Delta_\zeta\right)}
\right\rangle,\\
G &\approx& \left\langle
    \frac{\delta^{iso}}{C^0_\delta 
  - \frac{ (C^0_{\delta})_* }{ (P^0_{\delta})_* }P^0_\delta}
\right\rangle.
\end{eqnarray}
A detailed derivation of the above equations is provided in
Full Appendix C.2.

\section{Simulations and Observations}

\subsection{Simulated PSF Anisotropies and Corrections}
\label{subsec:sim}

We use simulations to test and assess the limitations of
our PSF correction scheme for the flexion measurement.
To do this, we assume particular models for describing the
isotropic/anisotropic PSF and the surface brightness distribution
for a source. In the present simulations 
observational noise and lensing effects are not taken into account.

First, we assume for 
the stellar surface brightness distribution
a two-dimensional Dirac delta function,
$I_*(\btheta)=\delta_D^2(\btheta)$,
while for the galaxy surface brightness distribution
a truncated Gaussian as
\begin{equation}
\label{eq:sim_gal}
I_{\rm gal}(\btheta) =
\exp\left(-\frac{|\btheta|^2}{2R_{\rm gal}^2}\right)
-
\exp\left(-\frac{R_{\rm max}^2}{2R_{\rm gal}^2}\right)
\ \ \ \ {\rm for} \ |\btheta|\le R_{\rm max},
\end{equation}
where $R_{\rm gal}$ and $R_{\rm max}$ 
are the Gaussian dispersion and the truncation radius
of $I_{\rm gal}$, respectively.
In the following we set $R_{\rm max}=3R_{\rm gal}$.
Next, we assume the isotropic part of PSF, $P^{iso}$, also follows
a truncated Gaussian of the form:
\begin{equation}
\label{eq:sim_iso}
P^{iso}(\btheta) = \frac{1}{2\pi\sigma_{iso}^2}\exp\left(
-\frac{|\btheta|^2}{2\sigma_{iso}^2}
\right)
-
\frac{1}{2\pi\sigma_{iso}^2}\exp\left(
-\frac{\theta_{\rm max}^2}{2\sigma_{iso}^2}
\right)  \ \ \ \ {\rm for} \ |\btheta|\le \theta_{\rm max},
\end{equation}
where $\sigma_{iso}$ and $\theta_{\rm max}$ are the Gaussian dispersion
and the truncation radius of $P^{iso}$, respectively.
In the following we set $\theta_{\rm max}=3\sigma_{iso}$.
Finally, we adopt the PSF anisotropy kernel $q(\btheta)$ of the
following form:
\begin{equation}
\label{eq:sim_aniso}
q(\btheta) = A_{\rm aniso}\frac{\theta_1}{|\btheta|^2}
\ \ \ \ {\rm for} \ |\btheta| \le \theta_{\rm aniso},
\end{equation}
where $A_{\rm aniso}$ is the normalization factor that controls the
strength of PSF anisotropy, and $\theta_{\rm aniso}$ is the truncation
radius. We have chosen the direction of anisotropy along the $x$-axis
($\theta_1$-axis).  Thus, the anisotropy kernel  $q(\btheta)$ has
two free parameters, $(A_{\rm aniso},\theta_{\rm aniso})$.
We vary the parameters $(A_{\rm aniso},\theta_{\rm aniso})$ to
test our anisotropic PSF correction scheme 
as a function of degree of PSF anisotropy. We take $\theta_{\rm aniso}$
in the range of $\theta_{\rm aniso}\in (0,\sigma_{iso})$.

Having set up the models, we then produce pixelized images for 
model stars and galaxies using the surface brightness distributions
$I_{*}(\btheta)$ and $I_{\rm gal}(\btheta)$, which are then
convolved with the model PSF to yield 
$I^{obs}_*(\btheta)$
and $I^{obs}_{\rm gal}(\btheta)$.
No observational noise or 
intrinsic/gravitational flexion has been added in the present
simulations. Here we consider the following set of Gaussian source
radii, $R_{\rm gal}=1\sigma_{iso}, 2\sigma_{iso}, 3\sigma_{iso}$.
For each set of  $(A_{\rm aniso},\theta_{\rm aniso})$, we measure 
various PSF moments, such as $D_q-\Delta_P\propto \zeta_q$,
from mock stellar images using the Gaussian weight
function of dispersion $r_g=\sigma_{iso}$. On the other hand,
we measure various shape moments for mock galaxy images of 
$I^{obs}_{\rm gal}(\btheta)$ 
using the Gaussian weight of 
$r_g=\sqrt{R_{\rm gal}^2+\sigma_{iso}^2}$ (i.e., Gaussian dispersion of
the PSF-convolved image). 
Then, 
we correct observed HOLICs $(\zeta^{obs},\delta^{obs})$
for the PSF
anisotropy to obtain $(\zeta^{iso}.\delta^{iso})$, which should
vanish
for a perfect PSF correction.

\begin{figure*}
\epsscale{1.0}
\plotone{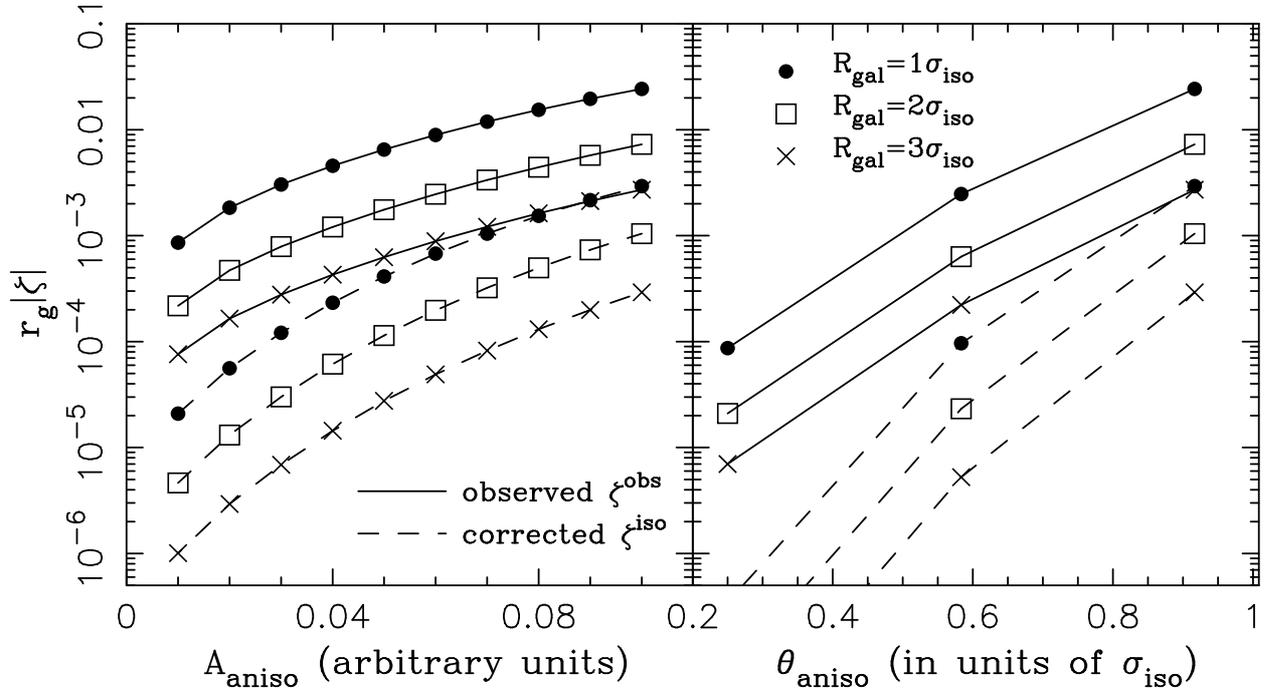}
\caption{
\label{fig:psfsim_zeta}
Test of anisotropic PSF correction for the first HOLICs of spin-1, 
$\zeta$, based on numerical simulations.
No observational noise and lensing signal is included.
The values of first HOLICs multiplied with the object detection radius,
$r_g|\zeta|$, are shown as a function of 
parameters $A_{\rm aniso}$ ({\it left}: $\theta_{\rm aniso}=0.92\sigma_{iso}$) 
and $\theta_{\rm aniso}$ ({\it right}: $A_{\rm aniso}=0.1$) 
for a model PSF anisotropy.
Solid lines indicate the observed values of first HOLICs ($\zeta^{obs}$)  
for Gaussian source images smeared with the model PSF,
and dashed lines indicate the residual values ($\zeta^{iso}$)
after correcting for the spin-1 PSF anisotropy.
No observational noise or lensing signal has been added.
The PSF consists of an isotropic part $P^{iso}$
described by a truncated Gaussian
with dispersion $\sigma_{iso}$ and an isotropic part
$q(\theta)=A_{aniso}\theta_1/|\btheta|^2$ 
truncated at $\theta=\theta_{aniso}$. 
Filled circles, open triangles, and crosses 
represent the measurements for a Gaussian source of 
dispersion $R_{\rm  gal}=1\sigma_{iso}, 2\sigma_{iso}, 3\sigma_{iso}$,
respectively. 
The $\zeta_{obs}$ is measured with a Gaussian weight function 
of dispersion  $r_g=\sqrt{R_{\rm gal}^2+\sigma_{iso}^2}$ 
from the surface brightness distribution smeared with the model PSF.
} 
\end{figure*}

\begin{figure*}
\epsscale{1.0}
\plotone{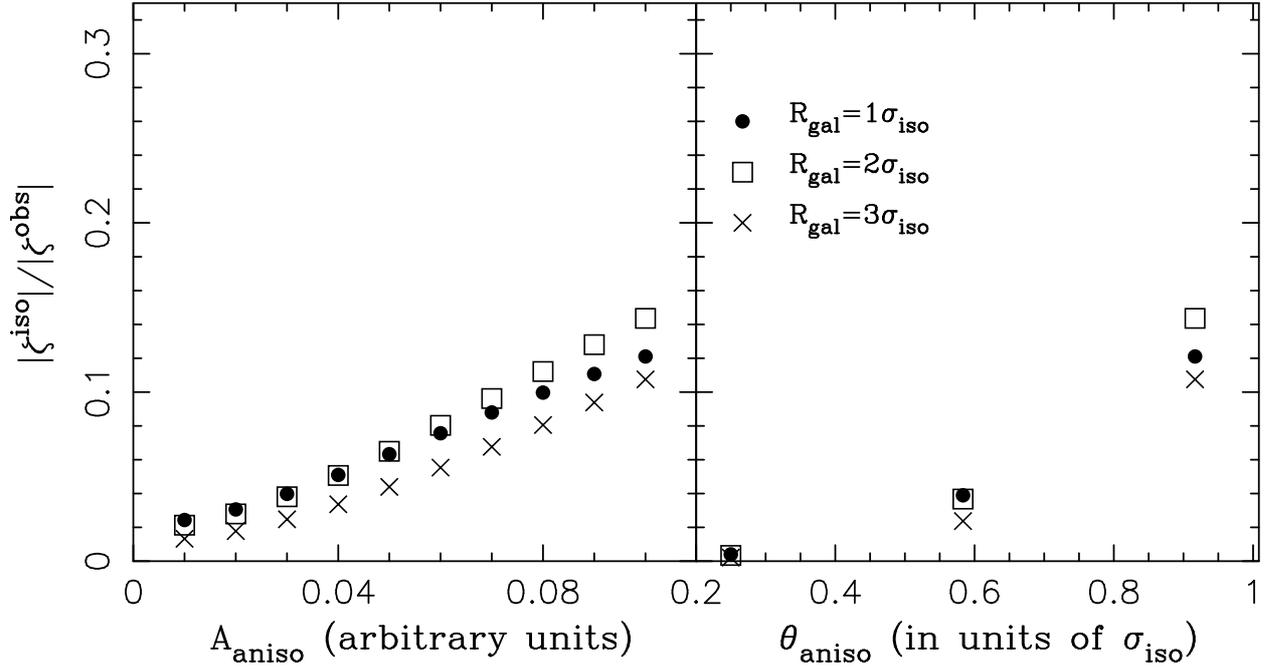}
\caption{
\label{fig:psfsim_zratio}
Ratio of residual to observed spin-1 PSF anisotropy, 
$|\zeta^{iso}|/|\zeta^{obs}|$, 
as a function of 
model parameters $A_{\rm aniso}$ ({\it left}) and $\theta_{\rm aniso}$ ({\it
 right}) for the PSF anisotropy kernel.
The PSF consists of an isotropic part $P^{iso}$
described by a truncated Gaussian
with dispersion $\sigma_{iso}$ and an isotropic part
$q(\theta)=A_{aniso}\theta_1/|\btheta|^2$ 
truncated at $\theta=\theta_{aniso}$. 
Filled circles, open triangles, and crosses 
represent the measurements for a Gaussian source of 
dispersion $R_{\rm  gal}=1\sigma_{iso}, 2\sigma_{iso}, 3\sigma_{iso}$,
respectively. 
} 
\end{figure*}

\begin{figure*}
\epsscale{1.0}
\plotone{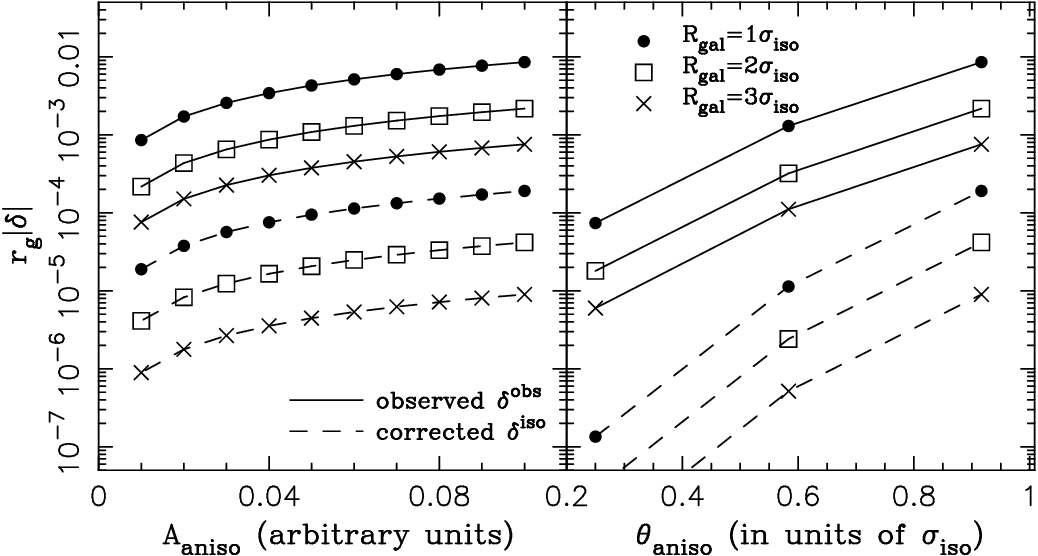}
\caption{
\label{fig:psfsim_delta}
Same as Figure 1 but for the second HOLICs, $\delta$, of spin-3.
} 
\end{figure*}

\begin{figure*}
\epsscale{1.0}
\plotone{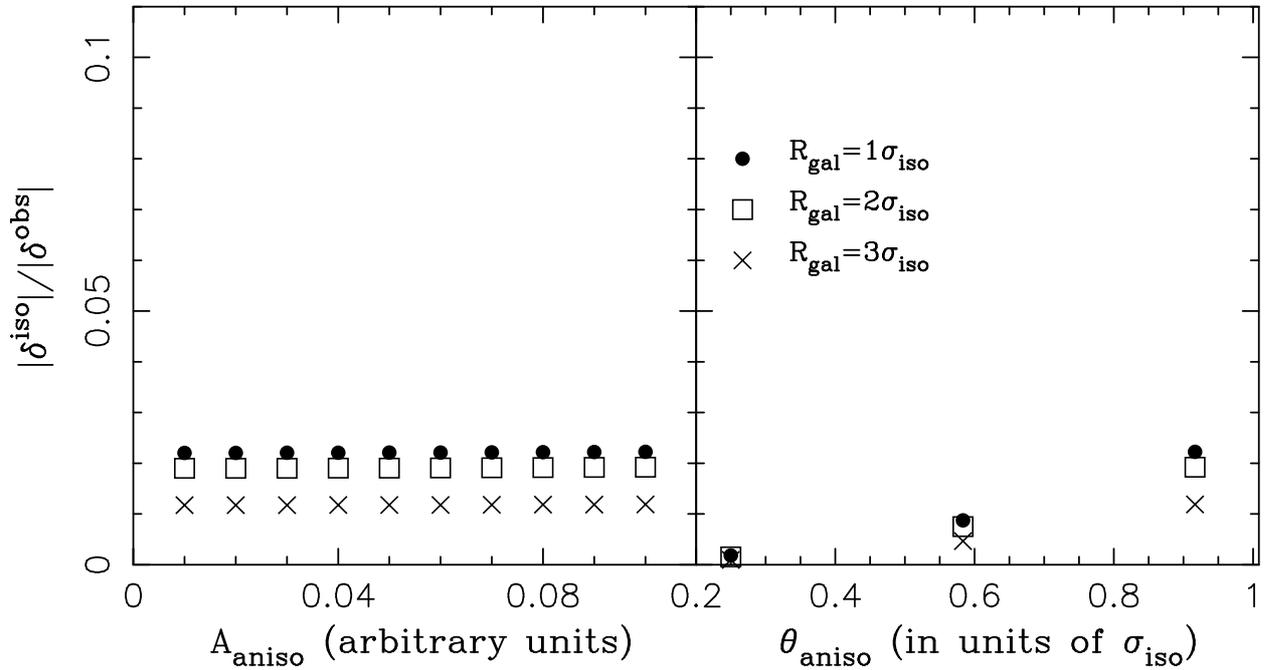}
\caption{
\label{fig:psfsim_dratio}
Same as Figure 2 but for the second HOLICs, $\delta$, of spin-3.
} 
\end{figure*}

We show in Figures \ref{fig:psfsim_zeta} and \ref{fig:psfsim_zratio}
results of anisotropic PSF correction for the first HOLICs $\zeta$ of
spin 1.
Figure \ref{fig:psfsim_zeta} shows the dimensionless 
$r_g|\zeta|$ before (solid) and after (dashed)
the anisotropic PSF correction 
as a function of degree of PSF anisotropy:
$A_{\rm aniso}$ (left panel) and $\theta_{\rm aniso}$ (right panel).
Firstly, it clearly shows that the spin-1 PSF anisotropy induced in 
mock galaxy images is considerably reduced after applying our PSF
correction method outlined in \S \ref{sec:ksb+}.
Secondly, we see a clear trend that the smaller the galaxy size,
the more severe the anisotropic PSF effect; or that $r_g|\zeta|$
increases with decreasing galaxy size, $R_{\rm gal}$.
In particular, when the source size is comparable to the size of PSF
($R_{\rm gal}=\sigma_{iso}$), the effect is larger 
about one order of magnitude than that of large sources with 
$R_{\rm gal}=3\sigma_{iso}$.
Figure \ref{fig:psfsim_zratio} shows 
the ratio 
$|\zeta^{iso}|/|\zeta^{obs}|$
of residual to observed PSF anisotropy in mock galaxy images.
We see that, overall, the fractional correction factor 
$|\zeta^{iso}|/|\zeta^{obs}|$ is larger for larger galaxy images.
Similar trends are also found for the second
HOLICs $\delta$ of spin 3, as shown in 
Figures \ref{fig:psfsim_delta} and \ref{fig:psfsim_dratio}.

\subsection{Flexion Analysis of Subaru A1689 Data}

We apply our flexion analysis method
based on the HOLICs moment approach to 
Subaru imaging observations of the cluster A1689. 
A1689 is 
a rich cluster of galaxies at a
moderately low redshift of $z=0.183$, 
having a large Einstein radius of $\approx 45$ arcsec 
($z_s\sim 1$; Broadhurst et al. 2005b).
A1689 is one of the best studied lensing clusters 
(e.g., 
Tyson \& Fisher 1995;
King, Clowe, \& Schneider et al. 2002;
Bardeau et al. 2005;
Broadhurst et al. 2005a;
Broadhurst et al. 2005b; 
Halkola et al. 2006;
Medezinski et al. 2007;
Leonard et al. 2007;
Limousin et al. 2007;
Umetsu, Broadhurst, Takada 2007; 
Umetsu \& Broadhurst 2007),
and therefore serves as an ideal target
for testing our flexion analysis pipeline.
Deep HST/ACS imaging of the central region of A1689
has revealed
$\sim 100$ multiply lensed images of $\sim 30$ background
galaxies (Broadhurst et al. 2005b), 
which allowed a detailed reconstruction of the mass
distribution in the cluster core
($10h^{-1} {\rm kpc} \simlt r \simlt 200 h^{-1}{\rm kpc}$).
Broadhurst et al. (2005a)
developed a method for
reconstructing the cluster mass profile 
by combining 
weak-lensing
tangential shear and magnification bias
measurements,
and derived a model-independent projected mass profile of 
the cluster out to its virial radius ($r\simlt 2 h^{-1}$ Mpc)
based on wide-field Subaru/Suprime-Cam data.
The combination of weak shear and magnification data breaks the
mass sheet degeneracy (Broadhurst, Taylor, \& Peacock 1995)
inherent in all reconstruction methods based
solely on the shape-distortion information 
(Schneider \& Seitz 1995).
Broadhurst et al. (2005a) found that 
the combined ACS and Subaru profile of the cluster 
is well fitted by an
NFW profile (Navarro, Frenk, \& Frenk 1997)
with 
a virial mass of $M_{\rm vir}=1.35\times 10^{15}h^{-1}M_{\odot}$
and 
a concentration of $c_{\rm vir} \sim 13.7$,
which is significantly larger than theoretically expected
($c_{\rm vir}\simeq 5$) for the standard LCDM model 
(Bullock et al. 2001; Neto et al. 2007).
Based on these Subaru data,
Umetsu \& Broadhurst (2007) used 
a maximum likelihood method
to reconstruct the two-dimensional mass map of A1689
from combined shear and magnification data, 
and found the azimuthally-averaged mass profile from the
full two-dimensional reconstruction is in good agreement with the
earlier results from the one-dimensional analysis by 
Broadhurst et al. (2005a), 
supporting the assumption of quasi-circular symmetry in the
projected mass distribution of A1689.
Recently, Leonard et al. (2007) performed a weak lensing analysis
of A1689 based on the ACS data 
by incorporating measurements of flexion as well as
weak shear and strong lensing, and their flexion reconstruction
has revealed mass substructures associated with 
small clumps of galaxies, while no flexion signal has been detected
at the cluster center, showing an under density around the location
of the cD galaxy.

For our flexion analysis of A1689,
we used Suprime-Cam $i'$-imaging data 
(Broadhurst et al. 2005a; Umetsu \& Broadhurst 2007), 
covering a field of $\sim 30'\times 25'$ with
$0\farcs 202 \,{\rm pixel}^{-1}$ sampling. 
The seeing FWHM in the co-added $i'$
image is $0\farcs 88$, and the limiting
magnitude is $i'=25.9$ for a $3\sigma$ detection within a $2''$ aperture
(see Broadhurst et al. 2005a). 
Since the flexion signal is weaker at larger angular scales 
(see OUF for detailed discussions), in the present flexion analysis
we discarded outer boundaries from the analysis and 
only used the central $3000\times 3000$ pixel region, 
corresponding an angular scale of $\approx 10'$ on a side,
or a physical scale of $1.3h^{-1}$ Mpc at the cluster redshift of
$z=0.183$.

\begin{figure*}
\epsscale{1.0} 
\plotone{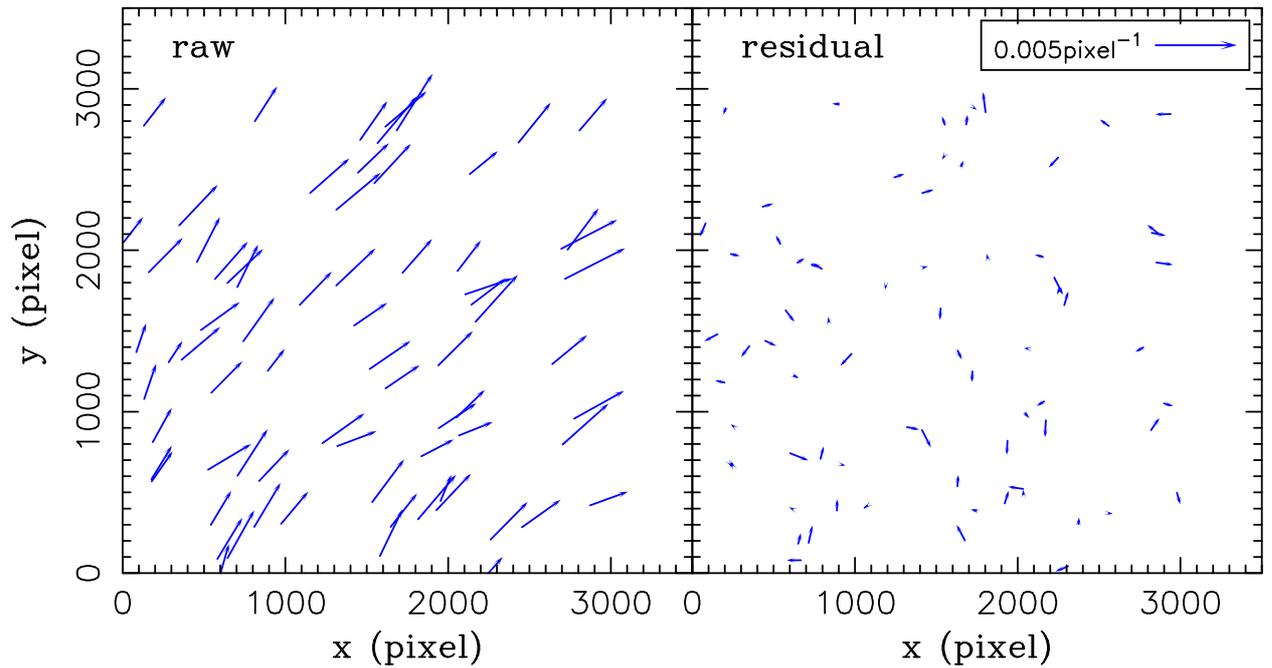}
\caption{
\label{fig:zetaq_field}
The spin-1 PSF anisotropy field $\zeta^{obs}_*(x,y)$
before ({\it left}) and after ({\it right}) the PSF correction
over the Subaru $i'$-band image of A1689. 
The spin-1 PSF anisotropy was measured from stellar shape moments
following the HOLICs formalism outlined in \S \ref{sec:ksb+}.
The orientation of the vectors shows the direction of the spin-1
 anisotropy, and the length is proportional to the magnitude of
 anisotropy.
A vector of $0.005 {\rm pixel}^{-1}$ is displayed in the inset panel.
} 
\end{figure*}

\begin{figure*}
\epsscale{1.0}
\plotone{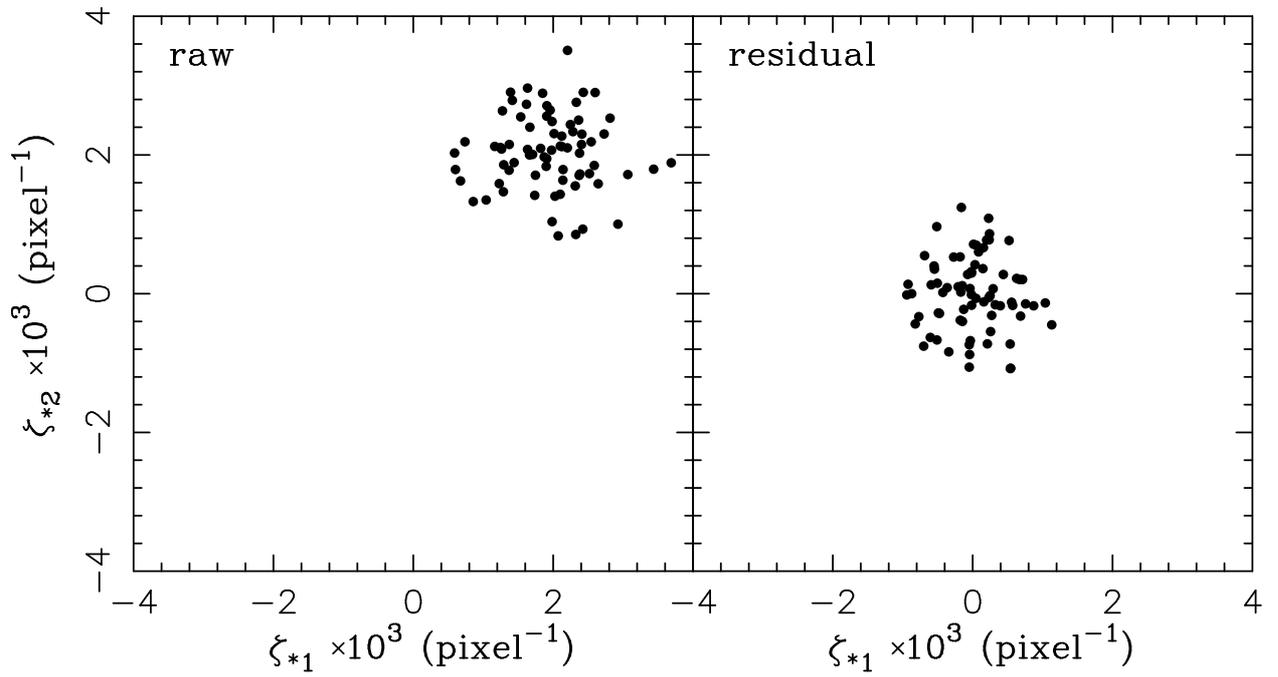}
\caption{
\label{fig:zetaq}
Comparison of spin-1 PSF anisotropy components before ({\it left})
and after ({\it right})
the PSF correction. 
}
\end{figure*}

We used our weak lensing analysis pipeline based on IMCAT (Kaiser et
al. 1995) extended to include our HOLICs moment method.
We selected a stellar sample of $N_*=73$ objects for measuring
the anisotropic/isotropic PSF effects.
On the basis of 
the simulation results, we excluded from our background galaxy sample
those small objects
whose half-light radius ($r_h$) and Gaussian detection radius ($r_g$)
are smaller than or comparable to
the PSF; here we selected galaxies with 
$0\farcs 6 < r_h < 2''$ and $r_g > 0\farcs 38$,
whereas the median values of stellar $r_h$ and $r_g$ are
$\langle r_{h*}\rangle \approx 0\farcs 48$ and 
$\langle r_{g*}\rangle \approx 0\farcs 34$ using $N_*=73$ stars.
These lower cutoffs in the galaxy size are essential 
for reliable flexion measurements,
because 
the smaller the object, 
the noisier its shape measurement due to pixelization noise;
and the shape of an image whose intrinsic size is smaller than
or comparable to the PSF size can be highly distorted and smeared,
as we have seen in \S \ref{subsec:sim}.
Faint objects will also yield noisy shape measurements, in particular
for the case of higher order shape moments. We thus selected bright
galaxies with $20 < i' < 25$ in the AB magnitude system.
Further we excluded from our background sample
those objects whose
flexion estimates are significantly larger than the model prediction 
($|F|\sim 0.1$ arcsec$^{-1}$)
using the best-fitting NFW profile 
derived from the joint Subaru and ACS analysis 
(Broadhurst et al. 2005a); 
with this model, 
we set an upper cutoff in flexion of $|F|<0.4$
arcsec$^{-1}$, and an upper cutoff in the first HOLICs of
$|\zeta^{iso}|<0.03$ arcsec$^{-1}$ (see equation [\ref{eq:HtoF}]). 
We note that the measurements of spin-3 HOLICs $\delta$ were found to 
be quite noisy (see OUF for detailed discussions), so that we discarded
the second flexion measurements from the present study.
Finally, these selection criteria yielded a sample of $791$ galaxies
usable for our flexion analysis, corresponding to a mean surface number
density of $\bar{n}_g= 7.75$ arcmin$^{-2}$.
Using the background galaxy sample above we found 
a mean value of 
$\langle F\rangle = 0.000223$ arcsec$^{-1}$ and a dispersion of
         $\sigma_F= 0.11245$ arcsec$^{-1}$.
In Figure \ref{fig:zetaq_field} we show the spatial distribution of
spin-1 PSF anisotropy as measured from stellar images before (left panel)
and after (right panel) the anisotropic PSF correction.
Figure \ref{fig:zetaq} compares the distribution of 
two components of complex spin-1 PSF anisotropy before (left panel) and
after(right panel) the anisotropic PSF correction.
As clearly shown in Figures \ref{fig:zetaq_field} and \ref{fig:zetaq},
the higher-order PSF anisotropy in observational data is indeed
significant, so that one needs to take into account the higher-order PSF
anisotropy correction in practical flexion measurements.

\begin{figure*} 
\epsscale{1.2}
\plotone{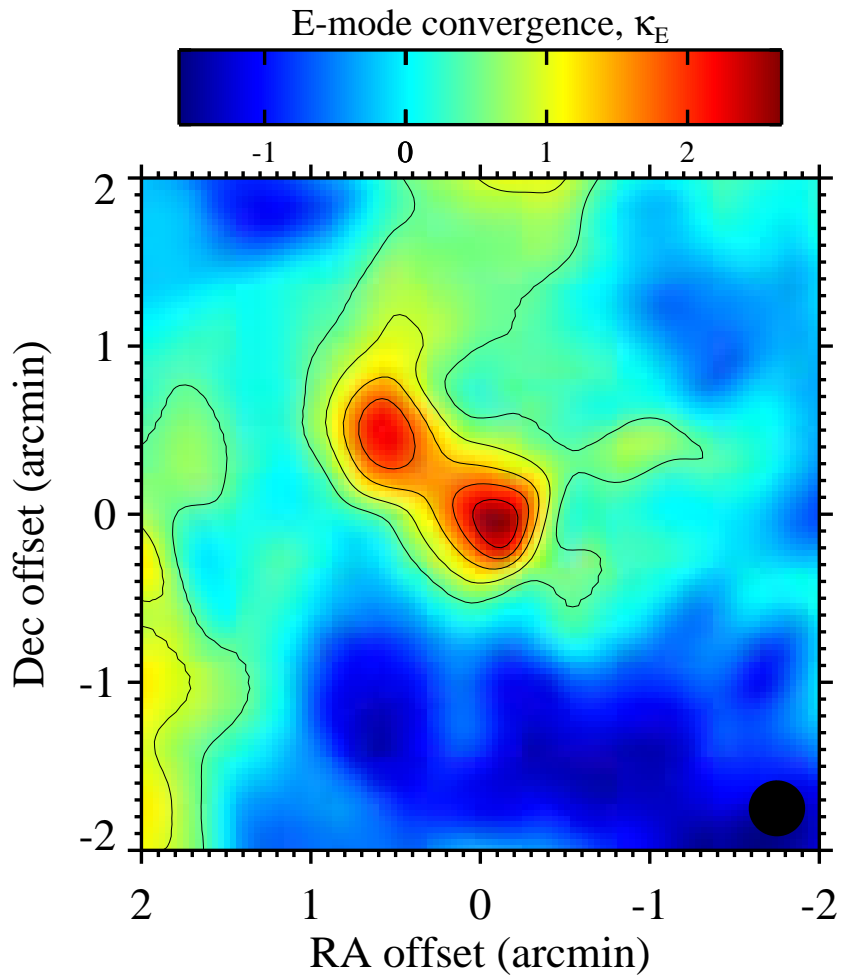}
\caption{
\label{fig:A1689_Emode}
The dimensionless surface mass density $\kappa$
of the galaxy cluster A1689 ($z=0.183$)
in the central $4'\times 4'$ region
reconstructed using the first flexion 
observed with Subaru telescope/Suprime-Cam.
The lowest contour and the contour interval are
at a $1\sigma$ level of
the reconstruction error ($\approx 0.51$)
estimated from the rms of the $B$-mode reconstruction.
The black, solid circle in the lower-right corner indicates the 
Gaussian FWHM ($=0\farcm 33$) used for the mass reconstruction.
}
\end{figure*}

\begin{figure*}
\epsscale{1.2}
\plotone{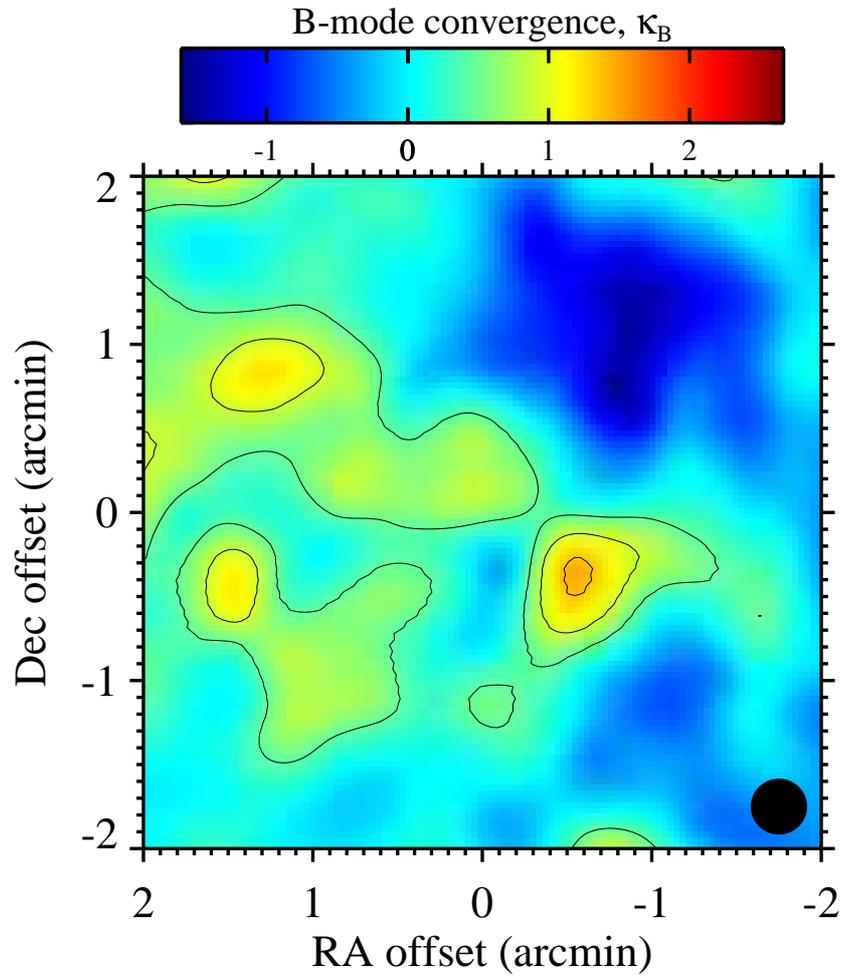}
\caption{
\label{fig:A1689_Bmode}
Same as Figure \ref{fig:A1689_Emode}
but for the
B-mode reconstruction
from the first flexion measured from Subaru data.
The color scale is the same as in Figure 
\ref{fig:A1689_Emode}.
The lowest contour and the contour interval are at a
$1\sigma$ level of the B-mode reconstruction 
($\sigma_B\approx 0.51$)
over the $8'\times 8'$ region.
}
\end{figure*}

\begin{figure*} 
\epsscale{1.0}
\plotone{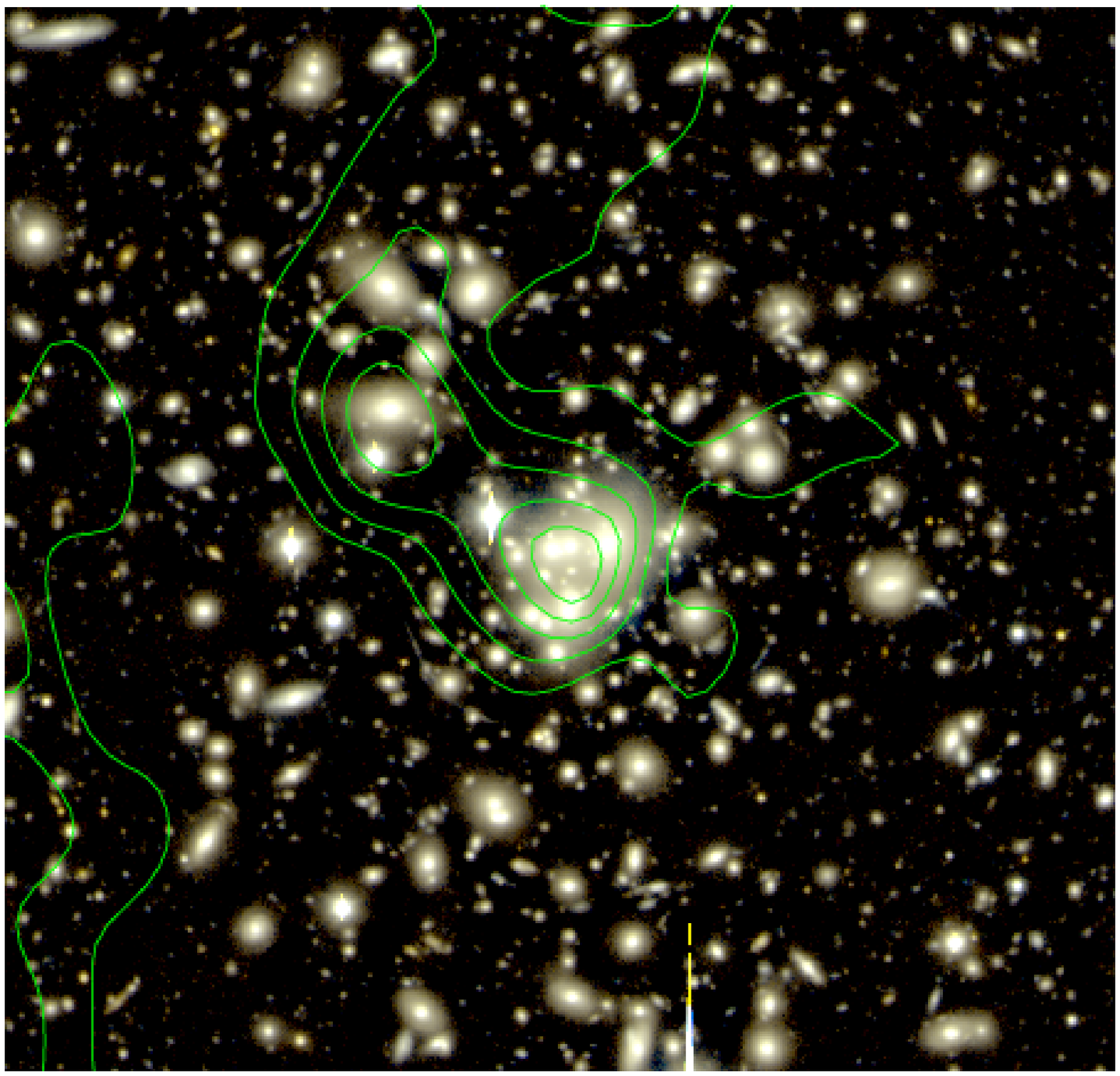}
\caption{
\label{fig:A1689_image}
False-color image of the central $4'\times 4'$ cluster region
composed of the Subaru/Suprime-Cam $V$ and $i'$ images.
Overlayed are contours of the lensing $\kappa$-field reconstructed from
 the first flexion measurements using 
the $i'$-band data.
The contours are spaced in units of $1\sigma (\approx 0.51)$ 
reconstruction error estimated from the rms of the $B$-mode
 reconstruction.
North is to the top, and East to the left.
} 
\end{figure*}

We use our first flexion measurements obtained with our moment-based
analysis method to reconstruct the projected mass distribution of
A1689. 
To do this we utilize
the Fourier-space relation between the first flexion $F$
and the lensing convergence $\kappa$ 
(\S 2.4 of OUF; Bacon et al. 2006) 
with the weak lensing approximation.
The field size for the mass reconstruction is
$9'\times 9'$, sampled with a grid of $256\times 256$ pixels,
over which the unconstrained $k=0$ mode is set to zero.
Figures \ref{fig:A1689_Emode} and \ref{fig:A1689_Bmode} show 
the $E$-mode convergence $\kappa=\kappa_E$ due to
lensing and the $B$-mode convergence $\kappa_B$ which is expected to
vanish in the weak lensing limit and can thus be used to monitor the
reconstruction error in the $E$-mode $\kappa$ map.
A central $8'\times 8'$ region is
displayed in Figures \ref{fig:A1689_Emode} and \ref{fig:A1689_Bmode}.
The reconstructed $\kappa$ maps were smoothed with a Gaussian filter
of ${\rm FWHM}=0\farcm 33$. 
Table \ref{tab:stats} lists basic statistics of the reconstructed
$E$- and $B$-mode $\kappa$ fields measured in the central $8'\times 8'$
field.
The rms dispersion in the Gaussian smoothed $B$-mode $\kappa$ map is
obtained as  
$\sigma_B\approx 0.51$. The maximum and minimum values 
in the $B$-mode convergence field are $1.52$ and $-1.61$,
corresponding to $2.9 \sigma$ and $-3.1\sigma$ fluctuations
(see Table \ref{tab:stats}).
Figure \ref{fig:A1689_Emode} reveals two significant mass concentrations
in the $E$-mode $\kappa$ map
associated with clumps of bright galaxies.
The first peak has a peak value of $\kappa_E=2.66$, and 
is detected at 
$5.2\sigma$ significance.
This first peak is associated with
the central concentration of bright cluster galaxies including 
the cD galaxy, as shown in Figure \ref{fig:A1689_image}.
This central mass concentration associated with the brightest cluster
galaxies was not detected in the earlier ACS flexion analysis by Leonard et
al. (2007).
The ACS/Subaru best-fitting NFW model predicts $\kappa(z_s=1)\approx
2.5$ at $\theta \sim 0\farcm 1$.
The second peak 
($4.4\sigma$ significance),
on the other hand, 
is located $\approx 0\farcm 9$ to the northeast direction,
and is associated with a
local clump of bright galaxies (see Figure \ref{fig:A1689_image}),
having a peak value of $\kappa_E=2.23$. 
This second mass peak has been detected in the earlier lensing studies
based on the high resolution HST/ACS data 
(e.g, Broadhurst et al. 2005b; Leonard et al. 2007),
and its likely bimodality in the central region 
has been discussed previous studies
(Miralda-Escude \& Babul 1995;
Halkola, Seitz, \& Pannella 2006;
Limousin et al 2007;
Saha, Williams, \& Ferreras 2007).
However, as compared to the Subaru weak lensing analysis by 
Umetsu \& Broadhurst (2007), 
the flexion-based mass reconstruction cannot recover
the global cluster structure on larger angular scales 
($\simgt$ a few arcmin), as demonstrated by OUF.

\section{Discussion and Conclusions}

In the present paper, we have developed a method for
weak lensing flexion analysis by fully extending the KSB method
to include the measurement of HOLICs (OUF). 
In particular, we take into
account explicitly the weight function in calculations of noisy
shape moments and the effects of spin-1 and spin-3 PSF anisotropies,
as well as isotropic PSF smearing, in the limit of 
weak lensing and
small  PSF anisotropy ($q$).
The higher order weak lensing effect induces a centroid shift in the
observed image of the background (Goldberg \& Bacon 2005; OUF; Goldberg
\& Leonard 2007). In weighted moment calculations, this will yield
in the flexion measurement 
additional correction terms (relevant to $W'(x), W''(x),
W'''(x)$) that must be taken into
account by properly expanding the weight function $W(x)$.
It is found that neglecting these additional terms originated from the
Taylor expansion of $W(x)$ yields the same result as obtained by
Goldberg \& Leonard (2007; see Appendix therein).
We extended the KSB formalism to include 
the higher-order 
isotropic and anisotropic PSF effects relevant to spin-1 and spin-3
HOLICs by following the 
prescription given by KSB and Bartelmann \& Schneider (2001), which
provides direct relations between 
the observable HOLICs and underlying flexion in the weak lensing limit.

We have implemented in our analysis pipeline
our flexion analysis algorithm based on the HOLICs
moment approach, and tested the reliability and limitation of our
PSF correction scheme using numerical simulations.
Our simulation results show that (i) 
after applying our PSF correction method
the PSF-induced anisotropies 
in HOLICs of mock galaxy images can be considerably reduced 
by a factor of $10$--$100$,
depending on the strength of PSF anisotropy, 
(ii) those small galaxies whose angular size is smaller than or
comparable to the size of PSF suffer from severe anisotropic PSF
effects,
and that (iii) there is an overall trend that
the fractional correction factor is larger for larger galaxy images.
Therefore, our simulation results 
support the reliability of our PSF-correction scheme and 
its practical implementation.

Based on the simulation results, we have applied our flexion analysis
pipeline to 
ground-based $i'$ imaging data of the rich cluster A1689 ($z=0.183$)
taken with Subaru/Suprime-Cam.
Our flexion analysis of Subaru A1689 data revealed
a non-negligible, significant effect of higher-order PSF anisotropy
induced in stellar images
(Figures \ref{fig:zetaq_field} and \ref{fig:zetaq}). 
It is therefore important in practical 
flexion measurements to quantify and correct for the higher-order
anisotropic PSF effects.

Our mass reconstruction from
the first-flexion measurements
shows two significant 
($>4\sigma$)
mass structures associated
with concentrations of bright galaxies in the central cluster region:
the first peak ($5.2\sigma$)
associated with the central concentration of bright
galaxies including the cD galaxy,
and the second peak ($4.4\sigma$) associated with a clump of bright
galaxies located $\sim 1'$ northeast of the cluster center.
This significant detection of the second peak
confirms 
earlier ACS results from the strong lensing analysis 
(Broadhurst et al. 2005b; Halkola et al. 2006; Leonard et al. 2007)  
and the combined strong lensing, weak shear,
and flexion analysis by Leonard et al. (2007).
The central mass peak, however,
 was not recovered in the earlier flexion analysis
by Leonard et al. (2007) based on HST/ACS data.
Leonard et al. (2007) attributed this to 
their relatively large reconstruction error at the cluster center,
although they have a very large number density of 
background galaxies,
$\bar{n}_g\approx 75$ arcmin$^{-2}$.
On the other hand,
owing to our conservative selection criteria for the background sample, 
the mean number density of background galaxies used for the present analysis
is $\bar{n}_g=7.75$ arcmin$^{-2}$, which is almost one order of
magnitude smaller than that of the ACS data, 
and is about $20\%-30\%$
of a typical number density of 
magnitude/size-selected background galaxies usable for the
quadrupole shape measurements 
in ground-based Subaru observations
($\bar n_g\sim 30-40 {\rm arcmin}^{-2}$).
However, we found that it is rather 
important to remove small/faint galaxy images and noisy outliers in
flexion measurements since they are likely to be affected by 
the residual PSF anisotropy and/or observational noise in the shape
measurement (\S \ref{subsec:sim}).  
Besides, the smaller the object, the larger the amplitude of 
intrinsic flexion contributions. Recall that
flexion and HOLICs have a dimension of length inverse: 
The response to flexion is size-dependent, and
the amplitude of intrinsic flexion is inversely proportional to
the object size. Indeed, we find that inclusion of smaller objects
results in a noisy reconstruction.  
Similar values of the background number density, $\bar n_g\simlt 10 {\rm
arcmin}^{-2}$,  have been used in
recent quadrupole weak lensing analyses based on Subaru observations
(e.g., Broadhurst et al. 2005a; Umetsu \& Broadhurst 2007; 
weak lensing cluster mass measurements of Okabe \& Umetsu 2008).
In their studies only objects redder than the cluster sequence
are selected in color-magnitude space
for their weak lensing analysis,
because such a red population is expected to comprise only background
galaxies ($\bar z_s\sim 0.9$; see, e.g., Medezinski et al. 2007), 
made redder by relatively large $k$-corrections and with
negligible contamination by cluster galaxies (Broadhurst et al. 2005a;
Medezinski et al. 2007). 
However, the smaller number of objects implies a coarser angular resolution 
in the map-making for achieving a proper signal-to-noise ratio
(e.g., per-pixel ${\rm S/N}\simgt 1$). With $\bar n_g\sim 10 {\rm
arcmin}^{-2}$ for cluster quadrupole weak lensing, 
typical angular resolutions are about $1-2$ arcmin 
(e.g., Gaussian FWHM, or boxcar width). 
On the other hand, flexion measures essentially 
the gradient of the tidal gravitational shear field
(i.e., $F,G\propto \phi(r)/r^3$), and hence is relatively sensitive
to small-scale structures.
Therefore, our successful reconstruction of the mass substructures 
with a small 
background density, $\bar n_g\sim 8 {\rm arcmin}^{-2}$, could be
attributed to the superior sensitivity of flexion to small scale
structures (see OUF for detailed discussions)
and the here-adopted selection criteria for a background
galaxy sample for weak lensing flexion analysis.

Finally, we emphasize that our HOLICs formalism here is different from
the earlier work by Goldberg \& Leonard (2007) in that
(1) additional correction terms for the centroid shift, relevant
to the derivatives of the weight function,
 have been included
and 
(2) the spin-1 and spin-3 PSF anisotropies, as well as the isotropic PSF
smearing, have been taken into account under the assumption of small PSF
anisotropy ($q[\btheta]$), as done in the KSB formalism.
Our flexion-based mass reconstruction of A1689 demonstrates the power of
the generalized flexion analysis techniques
for quantitative and accurate measurements of the weak gravitational
lensing effects.


\acknowledgments

We thank Masahiro Takada for valuable discussions.
We thank the anonymous referee for invaluable comments and suggestions.
The work is partially supported by the COE program at Tohoku University.
This work in part supported by 
the National Science Council of Taiwan
under the grant NSC95-2112-M-001-074-MY2.


\appendix

\section{HOLICs Formalism in the Complex Form}
\label{sec:basis}

\subsection{Complex Displacement}

In this Appendix, we adopt the following notations 
to represent complex quantities with 
the degree $N$ and the spin number $M$:
\begin{eqnarray}
X^N_M &\equiv& 
 (\dt_1 + i\dt_2)^{\frac{N+M}{2}}(\dt_1 +
 i\dt_2)^{*\frac{N-M}{2}}\nonumber\\
 &=& (\dt_1 + i\dt_2)^{\frac{N+M}{2}}(\dt_1 - i\dt_2)^{\frac{N-M}{2}},\\
Y^N_M &\equiv& 
 (\db_1 + i\db_2)^{\frac{N+M}{2}}(\db_1 + i\db_2)^{*\frac{N-M}{2}}\nonumber\\
&=& (\db_1 + i\db_2)^{\frac{N+M}{2}}(\db_1 - i\db_2)^{\frac{N-M}{2}}.
\end{eqnarray}
Here a quantity with the spin number of $-M$ 
is equivalent to the complex conjugate
of the corresponding spin-$M$ quantity.
Unless otherwise noted,
we shall use $X^N_M$ to represent complex quantities in 
the image plane, and $Y^N_M$ those in the source plane.

The product of $X^N_M$ and $X^L_K$ is expressed as
\begin{eqnarray}
X^N_MX^L_K 
   &=& X^{N+L}_{M+K},\nonumber\\
X^N_MX^{L*}_K 
   &=& X^{N+L}_{M-K} \ \ \ (M>K),\nonumber\\
  &=& (X^{N*}_MX^L_K)^* = \paren{X^{N+L}_{K-M}}^* \ \ \
   (M<K).
\end{eqnarray}
For arbitrary complex numbers $W$ and $Z$
the following identities hold:
\begin{eqnarray}
2\Real{WX^{N*}_M}X^L_K
&=&
\paren{WX^{N*}_M + W^*X^N_M}X^L_K 
= WX^{N+L}_{K-M} + W^*X^{N+L}_{M+K},\\
4\Real{WX^{N*}_M}\Real{ZX^{L*}_K}&=&
  \paren{WX^{N*}_M 
   + W^*X^N_M}\paren{ZX^{L*}_K + Z^*X^L_K},\nonumber\\
  &=& 
  WZ\paren{X^{N+L}_{K+M}}^* 
+ WZ^*X^{N+L}_{K-M} + W^*ZX^{N+L}_{-K+M} + W^*Z^*X^{N+L}_{K+M},\nonumber\\
&=&2\Real{WZ\paren{X^{N+L}_{K+M}}^*} + 2\Real{WZ^*X^{N+L}_{K-M}}.
\end{eqnarray}

\subsection{HOLICs Family}

We summarize in the complex form
a family of complex shape moments including HOLICs
relevant to the weak lensing flexion analysis.
 
\begin{eqnarray}
M&=\int d^2\theta I(\theta)W(X^2_0/\sigma^2)& {\rm(spin-0)}\nonumber\\
D&=\frac{\int d^2\theta I(\theta)W(X^2_0/\sigma^2) X^1_1}{M}
 &{\rm(spin-1)},\nonumber\\
\trQ&=\int d^2\theta I(\theta)W(X^2_0/\sigma^2) X^2_0 
 &{\rm(spin-0)},\nonumber\\
\chi&=\frac{\int d^2\theta I(\theta)W(X^2_0/\sigma^2) X^2_2}{\trQ}
 &{\rm(spin-2)},\nonumber\\
\zeta&=\frac{\int d^2\theta I(\theta)W(X^2_0/\sigma^2) X^3_1}{\xi}&{\rm(spin-1)},\nonumber\\
\delta&=\frac{\int d^2\theta I(\theta)W(X^2_0/\sigma^2) X^3_3}{\xi}&{\rm(spin-3)},\nonumber\\
\xi&={\int d^2\theta I(\theta)W(X^2_0/\sigma^2) X^4_0}&{\rm(spin-0)},\nonumber\\
\eta&=\frac{\int d^2\theta I(\theta)W(X^2_0/\sigma^2) X^4_2}{\xi}&{\rm(spin-2)},\nonumber\\
\lambda&=\frac{\int d^2\theta I(\theta)W(X^2_0/\sigma^2) X^4_4}{\xi}&{\rm(spin-4)},\nonumber\\
\iota_I&=\frac{\int d^2\theta I(\theta)W(X^2_0/\sigma^2) X^5_1}{\xi}&{\rm(spin-1)},\nonumber\\
\iotaIII&=\frac{\int d^2\theta I(\theta)W(X^2_0/\sigma^2) X^5_3}{\xi}&{\rm(spin-3)},\nonumber\\
\iota_V&=\frac{\int d^2\theta I(\theta)W(X^2_0/\sigma^2) X^5_5}{\xi}&{\rm(spin-5)},\nonumber\\
\upsilon_0&=\int d^2\theta I(\theta)W(X^2_0/\sigma^2) X^6_0&{\rm(spin-0)},\nonumber\\
\upsilonII&=\frac{\int d^2\theta I(\theta)W(X^2_0/\sigma^2) X^6_2}{\xi}&{\rm(spin-2)},\nonumber\\
\upsilonIV&=\frac{\int d^2\theta I(\theta)W(X^2_0/\sigma^2) X^6_4}{\xi}&{\rm(spin-4)},\nonumber\\
\upsilonVI&=\frac{\int d^2\theta I(\theta)W(X^2_0/\sigma^2) X^6_6}{\xi}&{\rm(spin-6)},\nonumber\\
\tau_I&=\frac{\int d^2\theta I(\theta)W(X^2_0/\sigma^2) X^7_1}{\xi}&{\rm(spin-1)},\nonumber\\
\tauIII&=\frac{\int d^2\theta I(\theta)W(X^2_0/\sigma^2) X^7_3}{\xi}&{\rm(spin-3)},\nonumber\\
\tau_V&=\frac{\int d^2\theta I(\theta)W(X^2_0/\sigma^2) X^7_5}{\xi}&{\rm(spin-5)},\nonumber\\
\tauVII&=\frac{\int d^2\theta I(\theta)W(X^2_0/\sigma^2) X^7_7}{\xi}&{\rm(spin-7)}.
\end{eqnarray}

\subsection{Differential Operators}


Let us first define the complex gradient operator $\partial$ as
\begin{equation}
\partial = \left(\frac{\partial}{\partial \theta_1} + i\frac{\partial}
{\partial \theta_2}\right).
\end{equation}

Operating $\partial$ on complex $X^N_M$ yields:
\begin{eqnarray}
\label{eq:GENdif}
&\hspace{-2cm}\partial X^1_1 &= \left(\frac{\partial}{\partial \theta_1} + i\frac{\partial}{\partial \theta_2}\right)(\theta_1 + i\theta_2)=(1-1)=0,\nonumber\\
&\hspace{-2cm}\partial^* X^1_1 
 &= \left(\frac{\partial}{\partial \theta_1} - i\frac{\partial}{\partial
				   \theta_2}\right)
 (\theta_1 +i\theta_2)=(1+1)=2,\nonumber\\
&\hspace{-2cm}\partial X^N_M 
 &= \partial\left( \left(X^1_1\right)^\frac{N+M}{2}\left(X^{1*}_1
		\right)^\frac{N-M}{2} \right) 
 = (N-M) \left(X^1_1\right)^{\frac{N+M}{2}}
 \left(X^{1*}_1\right)^\frac{N-M-2}{2} = (N-M)X^{N-1}_{M+1},\\
&\hspace{-2cm}\partial^* X^N_M 
 &= \partial^*\left( \left(X^1_1\right)^\frac{N+M}{2}
	       \left(X^{1*}_1\right)^\frac{N-M}{2} \right) 
 = (N+M) \left(X^1_1\right)^\frac{N+M-2}{2}\left(X^{1*}_1
\right)^{\frac{N-M}{2}} = (N+M)X^{N-1}_{M-1}.
\end{eqnarray}

Similarly, by operating $\partial$ on the weight function
$W(X_0^2/\sigma^2)$, one finds the following:
\begin{eqnarray}
\label{eq:GENdif2}
\partial W\left(\frac{X^2_0}{\sigma^2}\right) &=& \frac{2}{\sigma^2}X^1_1W'\left(\frac{X^2_0}{\sigma^2}\right),\nonumber\\
\partial^* W\left(\frac{X^2_0}{\sigma^2}\right) &=& \frac{2}{\sigma^2}X^{1*}_1W'\left(\frac{X^2_0}{\sigma^2}\right),
\end{eqnarray}
where $F$ and $G$ are the first and the second flexion defined as
$F=\cF/(1-\kappa)$ and $G=\cG/(1-\kappa)$, respectively.
Then, 
the complex displacement in the source plane, $\Delta\beta$,
is expressed in terms of the lensing convergence, shear and flexion as
\begin{equation}
\db = \db_1 +i\db_2 \equiv  Y^1_1 \approx (1-\kappa)
\left[
X^1_1 - gX^{1*}_1 - \frac{1}{4}
  \paren{2FX^2_0 + F^*X^2_2 + GX^{2*}_2}
\right].
\end{equation}
The integration measures in the source and image planes are related
in the following way
\begin{eqnarray}
d^2\beta &=&(1\tm\kappa)^2\Biggl(1 \tm 2\Real{FX^{1*}_1} \tm |g|^2 \tp \frac{1}{4}|F|^2X^2_0 \tm \frac{1}{4}|G|^2X^2_0\nonumber\\
&\tp& \frac{1}{2}\Real{F^2X^{2*}_2} \tm \Real{g^*FX^1_1} \tm \Real{g^*GX^{1*}_1} \tm \frac{1}{2}\Real{FG^*X^2_2}\Biggr)d^2\theta\nonumber\\
&\approx&(1-\kappa)^2\paren{1- 2\Real{FX^{1*}_1}}d^2\theta,
\end{eqnarray}
to the first order of reduced flexion (e.g. OUF).

\clearpage


\clearpage


\begin{deluxetable}{cllllll}
\tabletypesize{\scriptsize}
\tablecaption{
\label{tab:stats}
Basic statistics of the reconstructed $E$ and $B$ mode convergence fields}
\tablewidth{0pt}
\tablehead{
\colhead{field} &
\colhead{mean ($\bar\kappa$)} &
\colhead{$\sigma$}  &
\colhead{skewness \tablenotemark{a}}&
\colhead{kurtosis \tablenotemark{b}} &
\colhead{minimum}&
\colhead{maximum} 
} 
\startdata
$B$  & 0.058   &  0.512  &  $-0.334 \pm 0.2$  &  $0.043 \pm 0.4$ & -1.61 & 1.52\\
$E$  &-0.058   &  0.701  &  $-0.239 \pm 0.2$  &  $0.125 \pm 0.4$ & -2.09 & 2.66
\enddata
\tablecomments{The moments are calculated from 
the convergence within the central $8'\times 8'$ region.}
\tablenotetext{a}{Skewness defined as
 $\langle(\kappa-\bar\kappa)^3\rangle/\sigma^3$.}
\tablenotetext{b}{Kurtosis defined as
 $\langle(\kappa-\bar\kappa)^4 \rangle/\sigma^4-3$.}
\end{deluxetable}  




\begin{thebibliography}{99}


\bibitem{Bacon00}
	Bacon, D.~J., Refregier, A.~R., Ellis, R.~S. 2000, MNRAS, 318,
	625

\bibitem{Bacon06}
	Bacon, D. J.,
	Goldberg, D. M.,
	Rowe, B. T. P.,
	\& Taylor, A. N. 2006, \mnras, 365, 414

\bibitem{BARD1689}
	Bardeau, S. et al. 2005 A\&A, 434, 433 

\bibitem{review} 
	Bartelmann, M., \& Schneider, P. 2001, Phys.Rep., 340, 291
				       652,  937.

\bibitem{BTP}
	Broadhurst, T., Taylor, A.~N., \& 
	Peacock, J.~A. 1995, \apj, 438, 49


\bibitem{B05b} 
	Broadhurst, T. et al. 2005, \apj, 621, 53  (Broadhurst et al. 2005b)

\bibitem{B05a}
	Broadhurst, T., 
	Takada, M., 
	Umetsu, K.,
	Kong, X., 
	Arimoto, N., 
	Chiba, M.,  \& Futamase, T. 2005, 619,
	143L (Broadhurst et al. 2005a)


\bibitem{Bullock}
	Bullock, J.~S. et al. 2001, MNRAS, 321, 559


\bibitem{Erben00}
	Erben, T. et al. 2000, A\&A, 355, 23


\bibitem{GN02}
	Goldberg, D.~M. \& Natarajan, P. 2002,
	\apj, 564, 65


\bibitem{Flexion} 
	Goldberg, D. M., \& Bacon, D. J. 2005, ApJ, 619, 741

\bibitem{Flexion2} 
	Goldberg, D. M., \&  Leonard, A. 2006, ApJ, 660, 1003

\bibitem{halkola07}
	Halkola, A.
	Seitz, S.,
	\& Pannella, M. 2006, \mnras, 372, 1425

\bibitem{Hamana03} 
	Hamana, T. et al. 2003, ApJ, 597, 98


\bibitem{STEP1}
	Heymans, C. et al. 2006, MNRAS, 368, 1323 

\bibitem{Sextupole}
	Irwin, J., \& Shmakova, M. 2006, ApJ, 645, 17


\bibitem{KS93}
	Kaiser, N., \& Squires, G. 1993, ApJ, 404, 441

\bibitem{KSB} 
	Kaiser, N., Squires, G., Broadhurst, T. 1995, ApJ, 449, 460
	
\bibitem{KA1689} 
	King, L. J., Clowe, D. I., Schneider, P. 2002 A\&A, 383, 118 

\bibitem{GA1689} 
	Leonard, A., Goldberg, D.~M., Haaga, J.~L., 
	Massey, R. 2007, ApJ, 666, 51L

\bibitem{elinor}
        Medezinski, E., Broadhurst, T., Umetsu, K.,
        Coe, D., Benitez, N., Ford, H., Rephaeli, Y.,
        Arimoto, N., \& Kong, X. 2007, \apj, 663, 717


\bibitem{SW-A1689}
	Limousin, M. et al. 2007, ApJ, 668, 643


\bibitem{STEP2}
	Massey, R. et al. 2007, MNRAS, 376, 13

\bibitem{NFW}
	Navarro, J.~F., Frenk, C.~S., White, S.~D.~M., 1997, ApJ, 490, 493

\bibitem{Neto07}
	Neto, A.~F. et al. 2007, \mnras, 381, 1450


\bibitem{Okabe}
	Okabe, N. \& Umetsu, K. 2008, \pasj, in press 
	(arXiv:astro-ph/0702649)



\bibitem{HOLICs} 
	Okura, Y., Umetsu, K., \& Futamase, T. 2007, ApJ, 660, 995 

\bibitem{shapelet}
	Refregier, A. 2003, MNRAS, 338, 35

\bibitem{saha07}
	Saha, P., 
	Williams, L.~L.~R.,
	Ferreras, I. 2007, \apj, 663, 29

\bibitem{ss95}
	Schneider, P. \& Seitz, C. 1995, \aap 294, 411


\bibitem{Schneider07}
	Schneider, P. \& Er, X. 2007, submitted to A\&A (astro-ph/0709.1003)

\bibitem{TA1689}
	Tyson, J. A., \& Fisher, P., 1995 ApJ 446 L55

	
\bibitem{UTF1999}
	Umetsu, K, Tada, M., \& Futamase, T. 1999,
	Prog.~Theor.~Phys.~Suppl., 133, 53

\bibitem{UF2000}
	Umetsu, K, \& Futamase, T. 2000, ApJ, 539, L5

\bibitem{UA1689} 
	Umetsu, K., Takada, M., Broadhurst, T. 2007,
	Mod.~Phys.~Lett.~A, 22, 2099 (arXiv:astro-ph/0702096)

\bibitem{Umetsu07}
	Umetsu, K. \& Broadhurst, T. 2007, submitted to ApJ 
	(arXiv:astro-ph/0712.3441)

\bibitem{vanWaerbeke}
	Van Waerbeke et al. 2001, A\&A, 374, 757

\bibitem{Wittman}
Wittman, D. et al. 2001, ApJ, 557, 89L


\end{thebibliography}
\end{document}